\documentclass[a4paper,12pt]{article}
\usepackage[utf8]{inputenc}
\usepackage{cancel}
\usepackage{ulem}
\usepackage{amsfonts}
\usepackage{amssymb}
\usepackage{graphicx}
\usepackage{amsmath,bm}
\usepackage{enumerate}
\usepackage{mathtools}
\usepackage{tikz} \usetikzlibrary{calc}
\usepackage[countmax]{subfloat}
\usepackage{xcolor}
\usepackage{float}
\usepackage{color}
\usepackage{here}
\usepackage{cite}
\usepackage{subcaption}
\usepackage{mwe}
 
\usepackage{comment}
\usepackage[symbol]{footmisc}
\usepackage{mathrsfs}
\usepackage{float,epsfig}
\usepackage{dcolumn}% Align table columns on decimal point
\usepackage{pgfplots}
\usepackage{graphicx}% Include figure files
\usepackage{bm}% bold math
\usepackage{amsmath,amssymb,amsthm}
\usepackage{tikz-cd}
\usepackage[colorlinks=true,linkcolor=blue,citecolor=red]{hyperref}
\definecolor{lime}{HTML}{A6CE39}
\newcommand{\orcidicon}{%
	\begin{tikzpicture}
	\draw[lime, fill=lime] (0,0)
	circle [radius=0.16]
	node[white] {{\fontfamily{qag}\selectfont \tiny ID}};
	\draw[white, fill=white] (-0.0625,0.095)
	circle [radius=0.007];
	\end{tikzpicture}   \hspace{-2mm}
}
\newcommand\orcidHasan{{\href{https://orcid.org/0000-0001-7408-0910}{\orcidicon}}}
\newcommand\orcidKarima{{\href{https://orcid.org/0000-0001-5419-8516}{\orcidicon}}}
\newcommand\orcidFaical{{\href{https://orcid.org/0000-0002-2977-0821}{\orcidicon}}}
\title{\bf    On R\'enyi Microstructural Aspects of Asymptotically Flat Charged Black Holes: A Novel Duality}

\author{
	F. Barzi\orcidFaical\!\!$^{1,2}$\thanks{faical.barzi@edu.uiz.ac.ma},  
	H.  El Moumni\orcidHasan\!\!$^1$\thanks{h.elmoumni@uiz.ac.ma (Corresponding author)}, K. Masmar\orcidKarima\!\!$^{1}$\thanks{karima.masmar@gmail.com}\\
{\small $^{1}$ LPTHE, Physics Department, Faculty of Sciences, Ibnou Zohr University, Agadir, Morocco. }\\
{\small $^{2}$CRMEF, Regional Center for Education and Training Professions, Marrakesh, Morocco.}
}
\date{\today}
\begin{document} 
	\maketitle
\begin{abstract}
%We investigate the R\'enyi perspective on the microstructure of an asymptotically flat charged black hole. We propose a modification of the R\'enyi entropy to include compressibility effects and apply geometrothermodynamics within R\'enyi statistics. We probe the thermodynamics and statistical mechanics of the quantized asymptotically flat charged black hole within the Rényi entropy framework.

We explore the microstructure of asymptotically flat charged black holes through the lens of nonextensive R\'enyi statistics. A modified form of the R\'enyi entropy is proposed to incorporate compressibility effects, and geometrothermodynamics is applied within such a framework. Besides, by quantizing the black hole horizon area, we derive a microscopic description in terms of discrete degrees of freedom, with Rényi entropy providing a nonextensive generalization of the Bekenstein–Hawking entropy. We compute the Rényi partition function and probability distribution in the canonical ensemble, highlighting significant deviations from Boltzmann–Gibbs behavior at finite temperature due to nonextensive effects. The black hole undergoes a first-order Van der Waals–like phase transition between small and large configurations, characterized by discontinuities in entropy and latent heat, both of which are shown to depend on the number of area quanta. A critical threshold emerges beyond which the transition becomes athermal. Additionally, we establish a thermodynamic duality between charged-AdS black holes in Boltzmann–Gibbs statistics and charged-flat Rényi black holes, expressed via a conformal transformation of state variables. This duality extends naturally to higher dimensions. Our results provide new insights into the microstructure of black holes.

		{\noindent }
\end{abstract}
\newpage
	%\addcontentsline{toc}{section}{\nameref{appendix}}
	\tableofcontents
	
	%\newpage
	%\newpage

	%	 \tableofcontents

\section{Introduction}
Black holes have become crucial probes in the search for a unified theory of physics, as they sit at the intersection of general relativity, quantum field theory, and thermodynamics. These extreme conditions enable physicists to test the limits of classical gravitational theories and investigate quantum gravitational effects in a theoretical setting \cite{strominger1996microscopic, mathur2005fuzzball}. The study of black holes has revealed important principles, such as the holographic nature of gravity, as manifested in the AdS/CFT correspondence \cite{maldacena1999large}, as well as fundamental puzzles, including the black hole information paradox \cite{hawking1976breakdown}.

The development of black hole thermodynamics caused a radical paradigm shift in gravitational physics. Bekenstein proposed that black holes have entropy proportional to their horizon area \cite{bekenstein1973black}, and Hawking used quantum field theory in curved spacetime to derive black hole radiation and temperature \cite{hawking1975particle}. These insights led to the formulation of black hole analogs of the four laws of thermodynamics \cite{bardeen1973four, wald2001thermodynamics}, implying that black holes are thermodynamic objects with microphysical degrees of freedom. Traditional formulations rely on Boltzmann-Gibbs statistical mechanics, which violates the assumptions of additivity and extensivity of entropy in systems dominated by long-range interactions such as gravitation \cite{tsallis1988possible, padmanabhan2010thermodynamical}.

To address the shortcomings of the standard thermodynamic description, generalized entropy frameworks such as Tsallis and Rényi entropies have been proposed. These entropies apply the Boltzmann-Gibbs paradigm to nonextensive systems, where correlations and interactions span the entire system size. It maintains additivity for independent systems and allows controlled deviations from extensivity, making it a promising candidate for gravitational systems \cite{renyi1961measures, biro2013thermodynamics}. When applied to black holes, the Rényi formalism yields generalized thermodynamic laws that incorporate corrections to the Hawking temperature and horizon entropy, as well as richer phase structures, including Van der Waals-like transitions \cite{czinner2016renyi, czinner2017generalized, Promsiri:2020jga,promson2020renyi,Barzi2022,Barzi:2023mit,Barzi:2024vvo,Barzi2024yan,Barzi2024yin}.
Recent studies have explored Rényi entropy in the context of holography, which corresponds to entanglement entropies in dual conformal field theories. This reveals additional connections between gravitational thermodynamics and quantum information theory \cite{Dong2016re, Lewkowycz2013generalized}. These findings suggest that the thermodynamic behavior of black holes, particularly in AdS spacetimes, is inextricably linked to nonextensive statistical mechanics and quantum correlations, thereby bolstering the case for using generalized entropies, such as Rényi, to better model black hole microstates. 

It's commonly known that the thermal behavior of black holes strongly suggests the existence of an underlying microscopic structure that governs their macroscopic thermodynamic properties \cite{wei2015insight,KordZangeneh:2017lgs,base,Chabab:2017xdw,Miao:2017cyt}. A particularly insightful approach to this question comes from Ruppeiner geometry, which interprets thermodynamic fluctuations geometrically and offers a powerful lens through which to infer microscopic interactions solely from macroscopic thermodynamic data. Earlier investigations \cite{David:2002wn,Mirza:2007ev,Niu:2011tb,Dehyadegari:2016nkd} have reinforced this perspective, often employing tools from the AdS/CFT correspondence and string theory to relate black hole thermodynamics to dual quantum field theories \cite{Emparan:2006it,Horowitz:1996fn,Maldacena:1996gb}. More recently, Refs.\cite{base,Aounallah:2022rfo} present a quantized framework for charged AdS black holes, demonstrating that, under certain conditions, these black holes are unable to absorb discrete energy quanta. This behavior provides compelling evidence for the existence of a quantum microstructure underlying black hole thermodynamics, highlighting the role of quantization in constraining energy exchanges at the horizon scale.

The aim of this paper is to use nonextensive R\'enyi statistics to study various facets of the asymptotically flat charged and uncharged black hole microstructure. This includes modifying the R\'enyi statistics to better account for this possible microstructure. The microstructure and nature of the interactions between the black hole degrees of freedom are examined from a R\'enyi perspective, highlighting nonextensivity effects on black holes and a new and intriguing duality that links the asymptotically flat black hole in R\'enyi statistics to the asymptotically Anti-de-Sitter black hole under Boltzmaniann statistics.

The organization of the paper is as follows: In Sect.\ref{sect2} we propose a modified R\'enyi statistics by including an incompressibility parameter in the R\'enyi equation of state. We also apply geometrothermodynamics to the R\'enyi fluid to probe the interactions of its degrees of freedom. Section.\ref{sect3} is devoted to the quantization of the R\'enyi-Schwarzschild black hole horizon area and drawing the consequences of such procedure on the partition function, density of states, and the probability of the black hole phase in the canonical ensemble. Section.\ref{sect4} explores the equivalence between the Boltzmannian asymptotically-AdS black hole and the R\'enyian asymptotically-flat black hole through a conformal transformation which deepens their connection. A general discussion and closing remarks are given in Sect.\ref{sect5}.

\section{Rényi thermodynamics,  modified equation of state, and geometrothermodynamics}\label{sect2}

\subsection{Rényi thermodynamics and   modified Rényi equation of state}
\paragraph{}We begin our analysis from the first law of black hole thermodynamics in the context of Rényi entropy, which modifies the standard relation by incorporating nonextensive corrections. For a flat-charged Reissner–Nordström black hole, the first law and corresponding Smarr formula in the Rényi framework are given by \cite{Promsiri:2020jga,Barzi:2023mit}
\begin{equation}
dM = T_R\,dS_R + \Phi\,dQ, \qquad M = 2T_R S_R + \Phi Q. \label{eq:firstlaw}
\end{equation}
Here, $T_R$ and $S_R$ are the Rényi temperature and entropy, $M$ is the black hole mass, $Q$ is its charge, and $\Phi$ stands for the electric potential at the horizon. The Rényi entropy modifies the classical Bekenstein-Hawking result by introducing a nonextensive parameter $\lambda$, reflecting microscopic statistical deviations from the standard Gibbsian scheme. This entropy is expressed as
\begin{equation}
S_R = \frac{1}{\lambda} \ln(1 + \lambda S_{BH}), \label{eq:SR}
\end{equation}
where $S_{BH}=\pi r_h^2$ is the usual area law entropy, and $r_h$ is the event horizon radius.
Alfréd R\'enyi's formulation of entropy \cite{Renyi1959aqqq} provides a generalization of Boltzmann-Gibbs statistics that is compatible with thermodynamic equilibrium and the zeroth law

\begin{equation}\label{SR}
S_R=\frac{1}{\lambda}\ln\sum_ip^{1-\lambda}_i\; 
\xrightarrow{\; \lambda\rightarrow 0 \;}\;  S_{BG}=-\sum_ip_i\ln(p_i)
\end{equation}

 %Where $\lambda$ is the nonextensivity parameter and $p_i$ are the probabilities of the microstates accessible to the system. In the limit $\lambda\rightarrow 0$, the Boltzmann-Gibbs-Shannon entropy arises,
 %\begin{equation}
  % S_{BG}=-\sum_ip_i\ln(p_i).   
 %\end{equation}
 It is worth pointing out a common feature of the BG and R\'enyi entropies. In thermodynamic equilibrium, the probabilities are all equal, $p_i=p=\frac{1}{\Omega}$. In this case, the entropies read,
 \begin{eqnarray}\label{eq9}
     S_R=\ln(\Omega_R) \quad \text{and} \quad S_{BG}=\ln(\Omega_{BG})
 \end{eqnarray}
Such property results from the additivity of these entropies, although in the thermodynamic limit, BG entropy is extensive while R\'enyi is not. The R\'enyi entropy $S_R$ can also be understood as the formal logarithm of the Bekenstein-Hawking one $S_{BH}$ taken as the Tsallis entropy, $S_T$, such as shown in Eq.\eqref{eq:SR} and Refs. \cite{Czinner:2015eyk,Tannukij:2020njz,dilaton}. 

It is noteworthy that the R\'enyi entropy is given a formal additive composition law by the definition  Eq.\eqref{eq:SR}. Let $\Sigma_a$ and $\Sigma_b$ be two independent systems, we have for their joint R\'enyi entropy, using the pseudo-additivity composition rule of the Tsallis entropy\cite{tsallis1988possible,Abeintro},
\begin{align}\label{eq_additiv}
\begin{split}
S_R\left(\Sigma_a\cup \Sigma_b\right) =\,& \frac{1}{\lambda}\ln \biggl[ 1+ \lambda \biggl(S_T(\Sigma_a)+S_T(\Sigma_b)+\lambda S_T(\Sigma_a)S_T(\Sigma_b) \biggr) \biggr]  \\
=\,& \frac{1}{\lambda} \ln \biggl[ \biggl( 1+\lambda S_T(\Sigma_a) \biggr)\biggl( 1+\lambda S_T(\Sigma_b) \biggr) \biggr]  \\
=\,& \frac{1}{\lambda} \ln \biggl( 1+\lambda S_T(\Sigma_a) \biggr)+ \frac{1}{\lambda} \ln \biggl( 1+\lambda S_T(\Sigma_b) \biggr) \\
=\,& S_R(\Sigma_a)+S_R(\Sigma_b).
\end{split}
\end{align}
Consequently, Rényi entropy offers a practical method for dealing with black hole nonextensive characteristics while maintaining entropy additivity. When applied to black holes, through the generalized Euclidean path integral and the R\'enyi-Wick rotation introduced in \cite{Barzi:2024vvo}, one obtains the R\'enyi partition function,
\begin{equation}
    Z_\lambda[\beta]=\exp\left(\displaystyle -\frac{\beta^{2}}{16 \pi}-\frac{\lambda \beta^{4}}{512 \pi^{2}}\right).
\end{equation}
The R\'enyi free energy is then found to be $(\lambda\ll1)$,
\begin{align}
    F_R&=\displaystyle-\frac{1}{\beta}\ln\left(Z_\lambda\right)=\displaystyle \frac{\beta}{16 \pi }+\frac{\beta ^3 \lambda }{512 \pi ^2}+O\left(\lambda ^2\right),
\end{align}
The R\'enyi entropy is recovered to first order in $\lambda\ll1$ such as,
\begin{align}\label{eq_49d}
    S_R&=\beta^2\frac{\partial F_R}{\partial \beta}=S_{BH}-\frac{\lambda}{2}S_{BH}^2+O\left(\lambda ^2\right),
\end{align}

The Rényi entropy leads to the R\'enyi temperature:
\begin{equation}
T_R = \left(\frac{\partial S_R}{\partial M} \right)^{-1} = T_H (1 + \lambda S_{BH}), \label{eq:TR}
\end{equation}
where $T_H=\displaystyle\left(\frac{\partial S_{BH}}{\partial M}\right)^{-1}=\frac{r_h^2-Q^2}{4\pi r_h^3}$ is the Hawking temperature of the charged black hole. Consequently, the Rényi temperature becomes
\begin{equation}
T_R = \frac{(r_h^2 - Q^2)(1 + \lambda \pi r_h^2)}{4\pi r_h^3}. \label{eq:TR2}
\end{equation}
In the grand canonical ensemble, where the potential $\phi$ is held fixed, these quantities simplify to:
\begin{equation}
T_R = \frac{(1 - \Phi^2)(1 + \pi \lambda r_h^2)}{4\pi r_h}, \qquad M = \frac{r_h}{2}(1 + \Phi^2).
\end{equation}
These expressions encapsulate the leading-order corrections from Rényi nonextensive effects, under the assumption $0<\lambda<<1$, indicating small deviations from extensivity.

Following the extended thermodynamic framework, where the cosmological constant is interpreted as pressure, one can analogously define the Rényi pressure and associated thermodynamic volume:
\begin{equation}\label{eq_state}
P_R = \frac{3\lambda(r_h^2 - Q^2)}{32 r_h^2}, \qquad V = \frac{4\pi}{3} r_h^3.
\end{equation}
The specific volume, related to the microscopic degrees of freedom, is given by
\begin{equation}\label{spe_vol}
v = \frac{8l_p^2}{3} r_h = \frac{8}{3} r_h,
\end{equation}
in which Planck units $l_p=1$. The equation of state for the Rényi black hole fluid can then be derived as
\begin{equation}
P_R = \frac{T_R}{v} - \frac{2}{3\pi v^2} + \frac{128 Q^2}{27\pi v^4}. \label{eq:EOS}
\end{equation}
This relation reveals that the thermodynamics of charged-flat black holes under Rényi statistics exhibits a Van der Waals-like behavior, where the nonextensivity parameter $\lambda$ effectively mimics the role of the cosmological constant in standard Anti-de Sitter black hole thermodynamics. 
Next, we introduce the modified R\'enyi entropy to match the exact behavior of Van-der-Waals fluid. Indeed, the equation of state Eq.\eqref{eq:EOS}, for the charged-flat black hole, can be put in a form more inline with the Van-der-Waals equation of state, such as,
\begin{equation}\label{eq_state_modif}
\left(P_R+\frac{2}{3 \pi v^{2}} - \frac{128 Q^{2}}{27 \pi v^{4}}\right)\left(v-v_0\right)=T_R.
\end{equation}
Where $v_0$ is the co-volume encoding the non-ideal corrections due to interactions in the black hole microstructure. The equation of state Eq.\eqref{eq_state_modif}, with $v_0=0$, is a consequence of the mean potential energy per black hole "molecule" of the form,
\begin{equation}\label{modif_energy}
\epsilon=-\frac{a}{v}+\frac{a'}{v^3}.
\end{equation}
The first term corresponds to the standard Van der Waals mean-field (attractive) molecular interaction energy, while the second term arises from the Coulomb interaction. A direct comparison allows us to identify the corresponding coefficients $a$ and $a'$ as follows:
\begin{equation}\label{coeffs}
a=\frac{2}{3 \pi},\quad\text{   and }\quad a'=\frac{128Q^2}{81\pi}.
\end{equation}
Thus, the mean energy then reads 
\begin{equation}\label{eq21}
\epsilon=-\frac{2}{3\pi v}+\frac{128Q^2}{81\pi v^3}.
\end{equation}
The first term represents an attractive interaction energy that dominates at large distances, while the second term corresponds to a Coulombic repulsion that becomes significant at short ranges. In the context of Van der Waals fluids, the co-volume $v_0$ is nonzero and accounts for the finite size of the fluid constituents. This co-volume is also responsible for short-range repulsive interactions, which, in Eq.\eqref{eq21}, are modeled by the second term only. Accordingly, the emergence of Van der Waals–like behavior in both Rényi charged-flat black holes and charged-AdS black holes is fundamentally attributed to the presence of electric charge. To fully reproduce the thermodynamics of a Van der Waals fluid, a meaningful modification of the equation of state for the charged-flat black hole, Eq.\eqref{eq_state_modif}, has been proposed in the literature \cite{Rajagopal2014van,Ditta2023van}. This involves the inclusion of a non-vanishing co-volume $v_0$, which captures the effective size of the black hole’s microscopic degrees of freedom—often referred to as “black hole molecules”—such that
\begin{equation} 
\left(P_{RM}+\frac{2}{3 \pi v^{2}} - \frac{128 Q^{2}}{27 \pi v^{4}}\right)\left(v-v_0\right)=T_{RM}.
\end{equation}
Here, $P_{RM}$ and $T_{RM}$ denote the modified Rényi pressure and temperature, respectively. The first attempts to adjust the equation of state of black hole fluids within the extended phase space were introduced in \cite{Rajagopal2014van}, with further developments in more recent works \cite{Ditta2023van, Kaczmarek2024vann}. The core idea in these studies is to reproduce the exact thermodynamic behavior of a Van der Waals fluid, characterized by its two well-known parameters, which account for intermolecular attraction and finite molecular size, respectively. These approaches typically involve constructing new solutions to Einstein’s field equations that yield, in the extended phase space, a black hole equation of state explicitly matching that of the Van der Waals fluid. 
\textit{In contrast, the present study adopts a different perspective: the Van der Waals–like behavior is introduced directly through a modified Rényi entropy, which can then be applied to any existing black hole solution through an appropriate Wick rotation}\cite{Barzi:2024vvo}. A key advantage of this method is that it preserves the explicit spherical symmetry of the spacetime metric, while offering the flexibility to investigate a broad class of black hole solutions within a unified nonextensive thermodynamic framework.
% The mean potential energy per molecule becomes,
% \begin{eqnarray}
%     E=\int_{v_0}^{v}\left(\frac{2}{3\pi v^2}-\frac{128Q^2}{27\pi v^4}\right)d \overline{v}\\
%     =-\frac{2}{3\pi v}+\frac{128Q^2}{81\pi v^3}+\left(\frac{2}{3\pi v_0}-\frac{128Q^2}{81\pi v_0^3}\right)
% \end{eqnarray}
% \textcolor{blue}{integrate the pressure to find energy per molecule a generalization of eq 22}

Undoubtedly, since all degrees of freedom of the black hole live on its horizon, $v_0$ is proportional to the Planck's area $l_p^2$, and would therefore impose a lower bound on the horizon radius $r_h$ through Eq.\eqref{spe_vol}, such as,
\begin{equation} 
v>v_0 \implies r_h>r_{min},
\end{equation}
where $\displaystyle r_{min}= \frac{3}{8v_0 l_p^2 }$. To illustrate the implications of the modified equation of state, we begin with the simplest case: the Schwarzschild black hole ($Q=0$). In this scenario, the equation of state simplifies to
\begin{equation} 
\left(P_R+\frac{2}{3 \pi v^{2}} \right)\left(v-v_0\right)\implies P_R=  \frac{T_R}{v-v_0}  - \frac{2}{3 \pi v^{2}}.
\end{equation}
A further simplification can be made by retaining only first-order corrections in $v_0$, an approximation that holds in the regime of large specific volumes, $v\gg v_0$. Under this assumption, the modified Rényi pressure takes the form
\begin{equation} 
P_R=  \frac{T_R}{v}  - \frac{2}{3 \pi v^{2}}+\frac{v_0T_R}{v^2}+O\left(\frac{v^2_0}{v^2}\right).
\end{equation}
By inverting this relation, we obtain an expression for the Rényi temperature
\begin{equation}\label{eq27}
T_R=\displaystyle \frac{\pi  v^{2}P_R }{\pi \left(v_0 + v\right)}+ \frac{2}{3\pi \left(v_0 + v\right)}.
\end{equation}
We will use equation Eq.\eqref{eq27} to compute a modified R\'enyi entropy denoted as $S_{RM}$, for the Schwarzschild R\'enyi-flat black hole. According to the definition of thermodynamic temperature 
\begin{equation}
    \displaystyle T_{RM}=\frac{\partial M}{\partial S_{RM}},
\end{equation}
This allows us to express the modified Rényi entropy as
\begin{equation} 
 S_{RM}=\int\frac{1}{T_{RM}} \frac{dM}{dv}dv+S_c.
\end{equation}
where $M$ is the ADM mass of the black hole, and $S_c$ is an integration constant. 
After performing the integration, we obtain the explicit form of the modified Rényi entropy
\begin{equation}\label{mod_entropy}
S_{RM}=\displaystyle\frac{\log{\left(  1+\lambda\pi  r_{h}^{2} \right)}}{\lambda}+\displaystyle \frac{3 \sqrt{\pi} v_0 \arctan{\left(\sqrt{\lambda}\sqrt{\pi}  r_{h} \right)}}{4 \sqrt{\lambda}},
\end{equation}
where Eq.~\eqref{spe_vol} was used to express the specific volume $v$ in terms of the horizon radius $r_h$. 
In the limit $\lambda\to0$, the entropy expansion yields
\begin{align}
   S_{RM}=\displaystyle \pi  r_h^2 +\frac{3\pi v_0 r_h}{4} -\frac{\pi ^2 \lambda r_h^3 \left(2 r_h+v_0\right)}{4}  +O\left(\lambda ^{3/2}\right)
\end{align}
From this, we identify the corrected Boltzmann–Gibbs entropy as
\begin{align}\label{eq31}
S_{GBM}&=\displaystyle \pi r_{h}^{2}+\frac{3 \pi v_0 r_{h}}{4},\\
   &=S_{GB}+\frac{3 \pi v_0 r_{h}}{4}.
\end{align}
The correction in Eq.~\eqref{eq31} is valid in the large specific volume regime, which corresponds to large horizon radii, $r_h>>r_0$, where $\displaystyle r_0=\frac{3}{8l_p^2}v_0$. The co-volume $v_0$ represents the incompressible volume effectively occupied by the black hole "molecules."

Possible physical origins of this volume scale can be found in quantum considerations. For instance, the Heisenberg uncertainty principle (HUP) and its generalized forms (GUP) \cite{Valero2025gup,Bosso2023gup,Maggiore1993gup} suggests a minimal volume element in phase space, bounded from below by $\displaystyle \propto\hbar^3$. A connected origin is the possibility of black hole remanents\cite{Ong2025gup} where it is proposed that Hawking evaporation may halt, leaving small, compact, and stable objects. Moreover, loop quantum gravity (LQG) offers an alternative interpretation: LQG corrections to the asymptotically AdS Schwarzschild black hole introduce a minimal horizon radius \cite{Wang2024uiu}, which depends on the Barbero–Immirzi parameter \cite{Barbero1995uiu}.

 \paragraph{}It is important to emphasize that the modified entropy given in Eq.~\eqref{mod_entropy} is an increasing and convex function of the energy (or mass), a necessary condition for any physically consistent entropy function to ensure the existence of a well-defined, positive temperature. The corrected Gibbs-Boltzmann entropy can be recast into the compact form
\begin{equation} 
S_{GBM}=\displaystyle \pi r_{h}^{2}
\left(1+\frac{2r_0}{r_{h}}\right).
\end{equation}
For a more transparent mathematical expression of the modified entropy, we define the new parameter, $\mu=\displaystyle3\sqrt{\frac{\pi}{4}}v_0$, then,
\begin{equation}\label{eq33}
S_{RM}=\frac{1}{\lambda}\log{\left(  1+\lambda S_{GB} \right)}+\displaystyle \frac{\mu}{\sqrt{\lambda}} \arctan{\left(\sqrt{\lambda}\sqrt{S_{GB}} \right)}.
\end{equation}
The parameter $\mu$ can be interpreted as \textit{the incompressibility parameter} of the black hole fluid. In the limit of small $\lambda$ and $\mu$, we obtain,
\begin{equation}\label{eq34}
    S_{RM}= S_{\text{GB}}- \frac{\lambda}{2}S_{\text{GB}}^2+\mu  \sqrt{S_{\text{GB}}}+O\left(\lambda \mu\right)
\end{equation}
This expansion highlights three types of contributions: the leading extensive entropy $S_{GB}$, a negative nonextensivity quadratic correction proportional to $\lambda$, and a nonextensive square-root correction proportional to the incompressibility parameter $\mu$, which reflects microscopic repulsive effects associated with the black hole fluid's constituents' finite size. In other words, Eqs.~\eqref{eq33} and \eqref{eq34} capture a fundamental interplay between nonextensivity and incompressibility, characterized respectively by the parameters $\lambda$ and $\mu$. Since entropy quantifies the number of quantum degrees of freedom ($dof$) of the black hole, the nonextensive term tends to suppress the effective number of $dof$, whereas the incompressibility term contributes positively, enhancing the entropy.

\begin{figure}[!ht]
    \centering
    \includegraphics[scale=0.6]{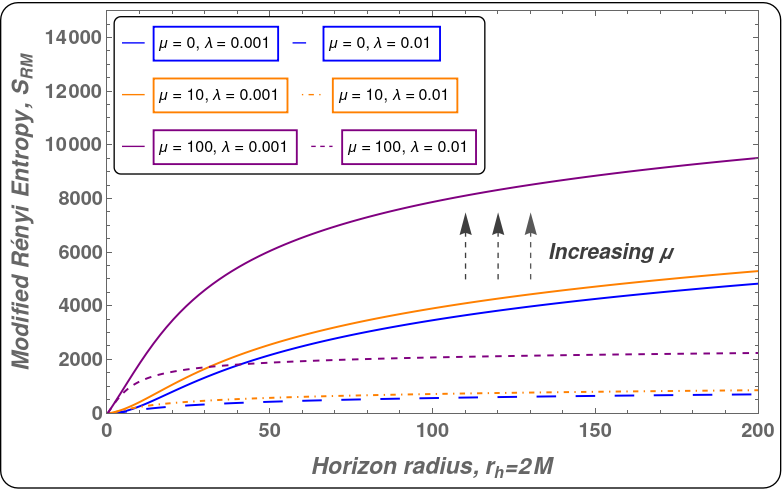}
    \vspace{-0.2cm}
    \caption{{\it \footnotesize Variations of the modified R\'enyi entropy in the $S_R-r_h$ plane with the nonextensivity and incompressibility ($\lambda$, $\mu$) parameters. Same color lines share the same incompressibility parameter, $\mu=0$ (Blue), $\mu=10$ (Violet), and $\mu=100$ (Red), while dashed and solid lines have nonextensivity parameters $\lambda=0.001$ and $\lambda=0.01$, respectively. The unmodified R\'enyi entropy corresponds to $\mu=0$.}} 
    \label{fig:fig2}
\end{figure}

\begin{figure}[!ht]
    \centering
    \includegraphics[scale=0.6]{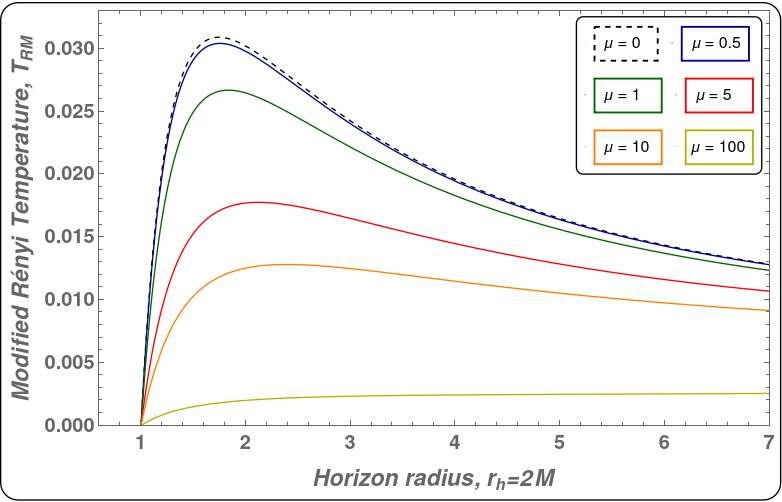}
    \vspace{-0.2cm}
    \caption{{\it \footnotesize Variations of the modified R\'enyi temperature in the $T_R-r_h$ plane with the nonextensivity and incompressibility ($\lambda$, $\mu$) parameters. Same color lines share the same incompressibility parameter, $\mu=0$ (Blue), $\mu=10$ (Violet), and $\mu=100$ (Red), while dashed and solid lines have nonextensivity parameters $\lambda=0.001$ and $\lambda=0.01$, respectively. The unmodified R\'enyi temperature corresponds to $\mu=0$.}} 
    \label{fig:temp}
\end{figure}

This competition leads to two distinct physical regimes:
\begin{itemize}
    \item {\bf Incompressibility-dominance regime}: at small entropy (or small black hole radius), the square-root term dominates, reflecting the prevalence of incompressibility effects;
    \item {\bf Nonextensivity-dominance regime}: at large entropy (or large black hole radius), the negative quadratic correction becomes dominant, indicating the suppression of $dof$ due to nonextensive statistics.
\end{itemize}
Besides, at an intermediate scale, where the two corrections become comparable, a threshold is established by setting
\begin{eqnarray}\label{eq35}
    \frac{\lambda}{2}S_{\text{GB}}^2=\mu  \sqrt{S_{\text{GB}}}\implies \displaystyle S_{\text{GB}}=\left(
    \frac{2\mu}{\lambda}\right)^{\frac{3}{2}},
\end{eqnarray}
This threshold entropy corresponds to a specific horizon radius $r_t$, given by
\begin{eqnarray}\label{eq36}
    r_t=\displaystyle\frac{1}{\sqrt{\pi}}\left(
    \frac{2\mu}{\lambda}\right)^{\frac{3}{4}}.
\end{eqnarray}
Therefore, the radius $r_t$ separates two regimes of black hole statistical behavior: 
    for $r_h<r_t$: the incompressibility-dominance regime, and  $r_h>r_t$ defines the nonextensivity-dominance scheme. Equivalently, the threshold mass is given by
\begin{equation}
\displaystyle M_t=\frac{1}{2\sqrt{\pi}}\left(
    \frac{2\mu}{\lambda}\right)^{\frac{3}{4}}.
\end{equation}
    \paragraph{} Figure~\ref{fig:fig2} illustrates the evolution of the modified Rényi entropy as a function of the nonextensivity parameter $\lambda$ and the incompressibility parameter $\mu$. 
    As expected, increasing $\mu$ shifts the entropy upward, reflecting the enhancing effect of incompressibility on the number of thermodynamic degrees of freedom. The R\'enyi modified temperature is computed as,
    \begin{equation}
        T_{RM}=\frac{\partial M}{\partial S_{RM}}\implies T_{RM}=\displaystyle\frac{\left(r_h^2-Q^2\right) \left(1+\pi  \lambda  r_h^2\right)}{2 \sqrt{\pi } \mu  r_h^2+4 \pi  r_h^3}.
    \end{equation}
    It is evident from the expression of the modified temperature that incompressibility effects reduce the black hole temperature. This trend is portrayed in Fig.\ref{fig:temp}. 
    
    In contrast to nonextensivity whose effects increase the temperature. The reduction is most prominent for small black holes, where the maxima of temperature are shifted downward.

\begin{figure}[!ht]
 \vspace{0cm}
		\centering
		\begin{tabbing}
		\hspace{-2.1cm}
		\includegraphics[scale=0.51]{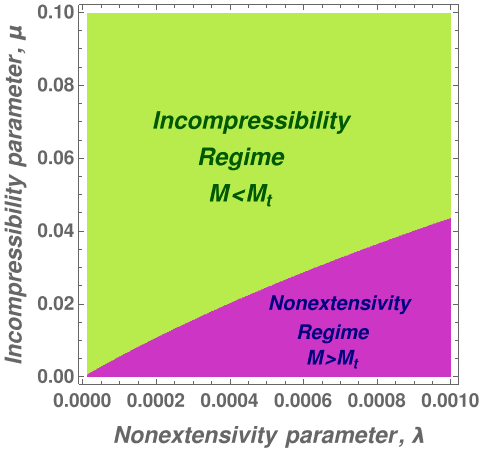}
		\hspace{-0.1cm}
		\includegraphics[scale=0.51]{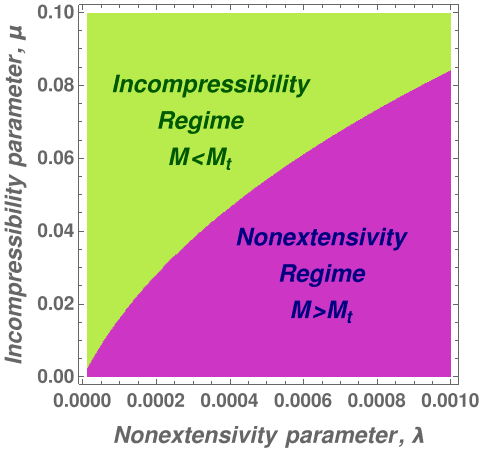}
            \hspace{-0.1cm}
		\includegraphics[scale=0.51]{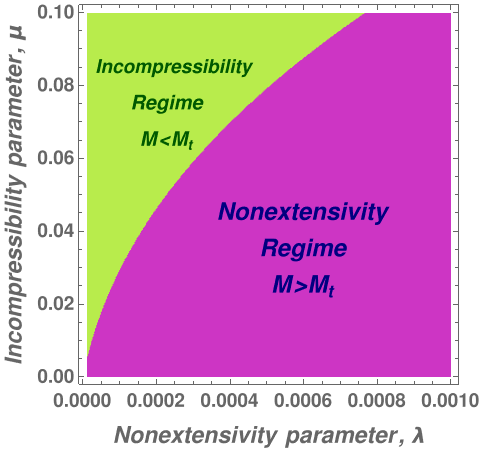}
		\end{tabbing}
        \vspace{-0.9cm}
        \begin{tabbing}
       \hspace{-2.1cm}
		\includegraphics[scale=0.51]{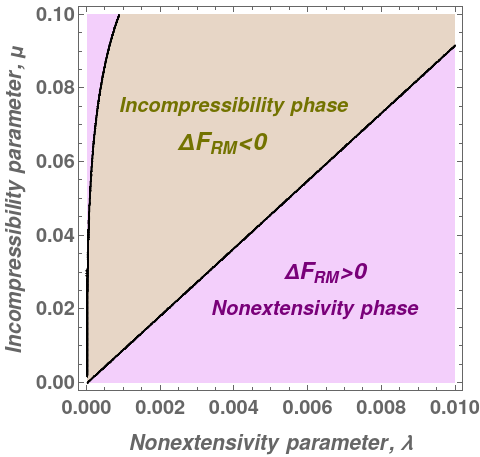}
		\hspace{-0.1cm}
		\includegraphics[scale=0.51]{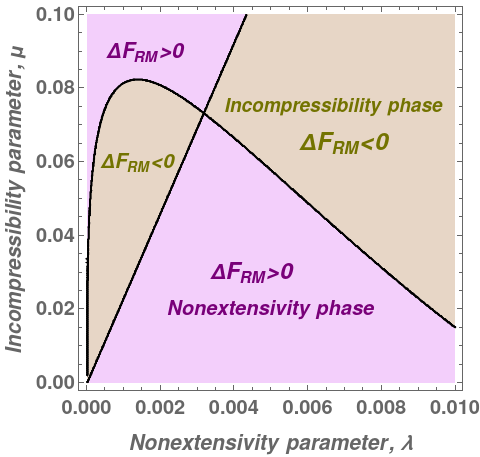}
            \hspace{-0.1cm}
		\includegraphics[scale=0.51]{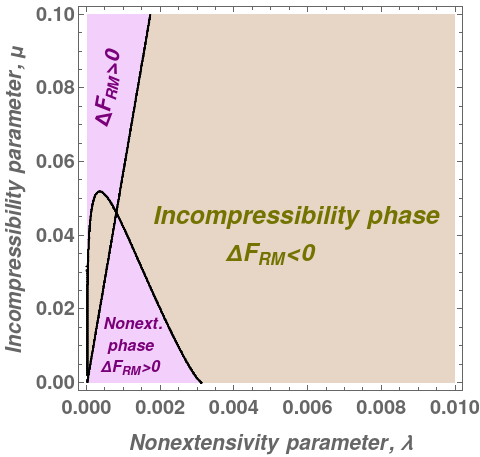}
        \end{tabbing}
		\vspace{-0.6cm}
		\caption{{\it\footnotesize \textbf{\textit{Top panel:}} Modified R\'enyi entropy of the uncharged flat black hole in the ($\lambda$, $\mu$)-plane. Green colored region indicates the dominance of the incompressibility effects. Magenta colored region indicates the dominance of nonextensivity effects. The black hole horizon radius $r_h$ (mass $\displaystyle M=\frac{r_h}{2}$) is increasing from left to right, $r_h=20,\, 50,\, 100$. \textbf{\textit{Bottom panel:}} The change in R\'enyi modified free energy of the uncharged flat black hole in the in the ($\lambda$, $\mu$)-plane. Beige region corresponds to a negative change in the free energy. Light-magenta region corresponds  to a positive change of the free energy. We took as reference free energy its value at the threshold mass $M_t$ and the horizon radius is increased from left to right to $r_h=5,\,10,\,20$.}}
		\label{fig:density_incompress}
	\end{figure}

We illustrated in Fig.\ref{fig:density_incompress} the interplay of the nonextensivity and incompressibility parameters $\lambda$ and $\mu$ and their influence on the entropy of the uncharged flat black hole, that is, on its number of microstates.

 It is uncovered that small black holes are dominated by incompressibility effects, while large black holes are dominated by nonextensivity effects. Depending on the parameters $\lambda$ and $\mu$, a given black hole of mass $M$ may display an incompressibility regime $(M<M_t)$ where its number of microscopic states is enhanced or the nonextensivity regime ($M>M_t$) where its number of microscopic states is suppressed. On the boundary between these regimes, a competition between these two regime ensues. 

Beyond simple regimes, one may expect a phase transition between a \textit{nonextensivity phase} and an \textit{incompressibility phase}, since entropy is a continuous function across the boundary as seen from Eq.\eqref{eq34}. This phase transition is continuous or second order, and thus no latent heat is generated. To visualize these phase transitions, we plotted \textit{in the bottom panel} of Fig.\ref{fig:density_incompress} the change in modified R\'enyi free energy taking as reference the free energy at the threshold mass $M_t$. In the nonextensivity phase, the entropy is suppressed, which increases the free energy of the black hole ($\Delta F_{RM}>0$). In contrast, in the incompressibility phase, the entropy is enhanced, reducing the free energy ($\Delta F_{RM}<0$). Future work will focus on a thorough examination of the phase profile and thermodynamic characteristics of the charged and uncharged flat black hole within the new modified R\'enyi statistics.

\paragraph{} We turn our attention in the next section to the investigation of the nature of the correlations between the microscopic constituents of the R\'enyi flat black hole through geometrothermodynamics.

%\newpage
\subsection{R\'enyi Geometrothermodynamics}
In this section, we examine the asymptotically flat R\'enyi charged black hole's thermodynamic geometry by investigating various forms of interactions between two molecules in the black hole fluid.  Fortunately, a powerful tool is provided by well-known thermodynamic geometries derived from the theory of thermodynamic fluctuations. We employ the geometrothermodynamics method based on Ruppeiner \cite{Ruppeiner1979eee,Ruppeiner1995ttt} and 
 Quevedo geometries \cite{Quevedo:2006xk,Quevedo:2007mj,Quevedo:2010tz}. It is worth pointing that historically the development of geometrothermodynamics has seen two distinct routes, the first one was an improvised introduction of a metric on the space of thermodynamic states, denoted $\mathbb{E}$, by Weinhold\cite{Weinhold1975geo} and Ruppeiner\cite{Ruppeiner1979eee}, the second, a more mathematically consistent approach\cite{Hermann1973geo, Mrugaa1978geo,Quevedo2003geo},  based on contact geometry of the thermodynamic phase space denoted $\mathbb{T}$, relies heavily on the additivity and extensivity of the coordinates of the two spaces $\mathbb{E}$ and $\mathbb{T}$ which include the entropy $S$, taken to be the Gibbs-Boltzmann entropy as it corresponds to most ordinary material systems, with short range interactions and local correlations. However, for black holes, the Bekenstein-Hawking entropy is non-additive and nonextensive, which may call in question the whole mathematical framework of geometrothermodynamics when applied to black holes, unless a suitable choice of entropy is made. Advantageously, the R\'enyi entropy exhibits a remarkable feature: it incorporates the nonextensivity of black holes while preserving its additivity, Eq.\eqref{eq_additiv}. From this point of view, R\'enyi entropy is better suited to play the role of entropy when considering the applicability of geometrothermodynamics to black holes.

\subsubsection{R\'enyi-Ruppeiner Geometrothermodynamics}
\begin{figure}[!ht]
	\centering
	\includegraphics[scale=0.65]{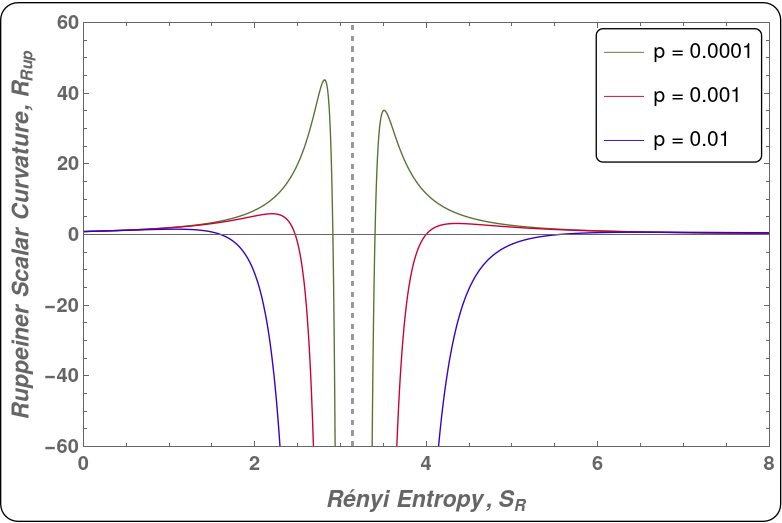}
	\caption{{\it\footnotesize Variations of Ruppeiner scalar curvature with R\'enyi entropy for different R\'enyi pressure. The vertical dashed line indicates a singularity at the extremal entropy $S_0=\pi Q^2 $. Electric charge is fixed at $Q=1$.}}
	\label{fig:fig6}
\end{figure}

\paragraph{}For the charged-flat R\'enyi black hole, the line element of the Ruppeiner geometry is defined in parameter space as,
\begin{equation}\label{eq41u}
    ds^2=\frac{1}{T_R}\frac{\partial^2 M}{\partial x^\mu\partial x^\nu}dx^\mu dx^\nu,
\end{equation}
where $x^\mu=\{S_R,\,Q\}$. Explicitly, it reads,
\begin{equation}\label{eq41uu}
    ds^2=\frac{1}{T_R}\left(\frac{\partial^2 M}{\partial S_R^2}dS_R^2+2\frac{\partial^2 M}{\partial S_R\partial Q}dS_R dQ+\frac{\partial^2 M}{\partial Q^2}dQ^2\right),
\end{equation}
The variations $dx$ in Eq.\eqref{eq41u} measure the fluctuations of the quantity $x$ from an equilibrium value $x_0$ for which the metric vanishes $ds^2=0$. Thus, the metric $ds^2$ is a measure of the likelihood of a fluctuation characterized by its distance from the equilibrium state. \textit{That is, the larger a fluctuation is, the farther it is from equilibrium and the less likely it is to manifest}. The Ruppeiner metric yields a thermodynamic scalar curvature $R_{Rup}$ in analogy with General Relativity (GR). The sign of $R_{Rup}$ bears explicit information about the nature of intermolecular interaction within the black hole fluid. A positive $R_{Rup}>0$ ($R_{Rup}<0$) indicates repulsive (attractive) interactions, while a vanishing $R_{Rup}=0$ signifies the absence of any interactions at a distance among the molecules \cite{Belhaj:2015uwa,Chabab:2015ytz}. In Fig.\ref{fig:fig6} we plotted the variations of the Ruppeiner scalar curvature of the charged-flat R\'enyi black hole against the R\'enyi entropy. We consider the entropy $S_R$ and pressure $p$ as the relevant thermodynamic fluctuation coordinates and operate in the canonical fixed charge ensemble.

\paragraph{}The scalar curvature changes sign twice at pressure-dependent values of the entropy. Dominant repulsive interactions in the black hole fluid are revealed below and above this interval by the positive sign of $ R_{Rup}$ with an asymptotic behavior toward zero for large entropies and approaching a small but finite value, $ R_{Rup}\rightarrow\frac{3}{\pi Q^2}$, for very small entropies, whereas prevailing attractive interactions are indicated by the negative sign within it. Moreover, a substantial change in the scalar curvature is observed around $S_0=\pi Q^2$, indicating an abrupt intensification of the interactions among the black hole "molecules" at this state. The singularity at $S_0$ is a consequence of the extremality of the black hole since this value corresponds to $M=Q$ and a naked singularity. From the prevailing interpretation of the Ruppeiner curvature, the vanishing of the curvature at large entropies indicates an ideal gas behavior revealed by the absence of interactions ($ R_{Rup}=0$) amid the microscopic degrees of freedom. Additionally, the zeros of the scalar metric denote a potential phase transition from large black holes (LBH) to small ones, where the interactions assure the reorganization of the microscopic constituents into the new phase. Subsequently, the intensity of these interactions reduces to a smaller but positive value as the black hole enters the Planckian regime. 

\begin{figure}[!ht]
 \vspace{0cm}
		\centering
		\begin{tabbing}
		\hspace{-2.1cm}
		\includegraphics[scale=0.51]{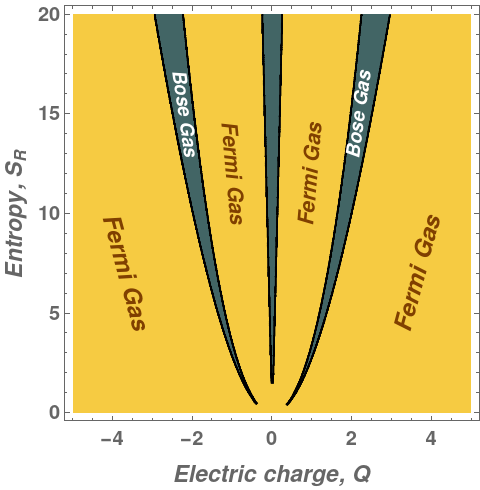}
		\hspace{-0.1cm}
		\includegraphics[scale=0.51]{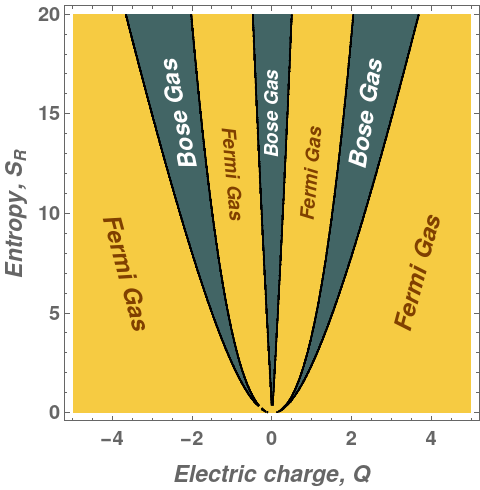}
            \hspace{-0.1cm}
		\includegraphics[scale=0.51]{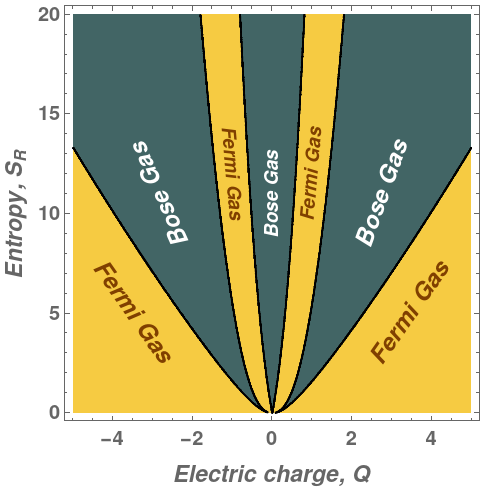}
		\end{tabbing}
		\vspace{-0.9cm}
              \vspace{0cm}
		\centering
		\begin{tabbing}
		\hspace{-1.9 cm}
		\includegraphics[scale=0.51]{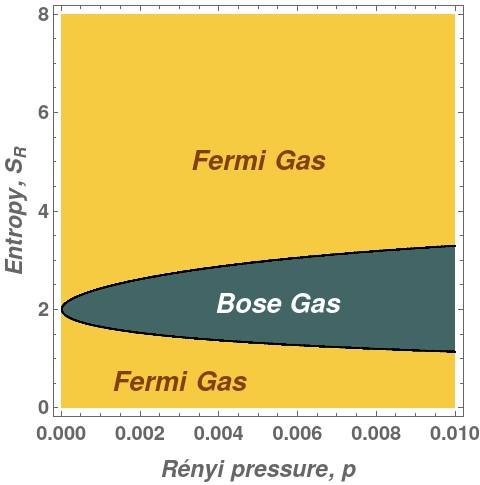}
		\hspace{-0.1cm}
		\includegraphics[scale=0.51]{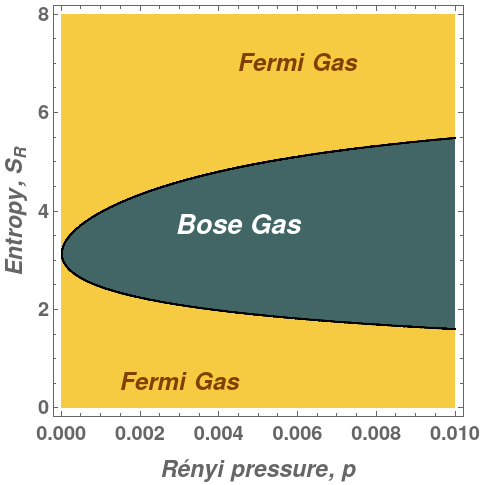}
            \hspace{-0.1cm}
		\includegraphics[scale=0.51]{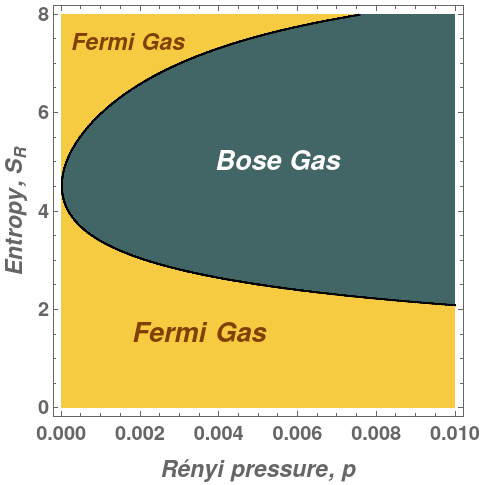}
		\end{tabbing}
		\vspace{-0.6cm}
		\caption{{\it\footnotesize \textit{\textbf{Top panel:}} The Ruppeiner scalar curvature as a function of the R\'enyi entropy $S_R$ and the electric charge $Q$. \textbf{The blue region }indicates a bosonic gaz behavior $R_{Rup}<0$, \textbf{the yellow region }shows a fermionic gaz behavior $R_{Rup}>0$ and \textbf{the black line} denotes an ideal gaz comportment $R_{Rup}=0$. R\'enyi Pressure is increasing from left to right: $p=0.0002,\,0.001,\,0.01$. \textit{\textbf{Bottom panel:}} The Ruppeiner scalar curvature as a function of the R\'enyi entropy $S_R$ and the R\'enyi pressure $p$. \textbf{The blue region }indicates a bosonic gaz behavior $R_{Rup}<0$, \textbf{the yellow region }shows a fermionic gaz behavior $R_{Rup}>0$ and \textbf{the black line} denotes an ideal gaz comportment $R_{Rup}=0$. Electric charge is increasing from left to right: $Q=0.8,\,1,\,1.2$.}}
		\label{fig:density_rup}
	\end{figure}
Fig.\ref{fig:density_rup} renders the Ruppeiner scalar curvature, $R_{Rup}$, in the $(Q,S_R)$ and $(p,S_R)$ planes. It is unveiled that the black hole fluid exhibits three distinct behaviors according to the sign of $R_{Rup}$\cite{Oshima1999geo,wei2015insight}, a fermionic gas-like depicted by the yellow region ($R_{Rup}>0$), a bosonic gas-like illustrated by the blue region and an ideal gas-like at the black boundary separating the two colored regions. Furthermore, as the R\'enyi pressure of the black hole fluid is increased (top panel), bosonic behavior dominates over the fermionic one demonstrated by the relative areas occupied by each one of them. The pressure being a measure of the nonextensivity of the black hole as defined by Eq.\eqref{eq_state}, nonextensivity tends to favor a bosonic nature for the microscopic degrees of freedom of the black hole. An identical trend is observed when the electric charge is increased (bottom panel), as the bosonic behavior becomes progressively more prominent.

\subsubsection{R\'enyi-Quevedo Geometrothermodynamics}
\begin{figure}[!ht]
	\centering
	\includegraphics[scale=0.65]{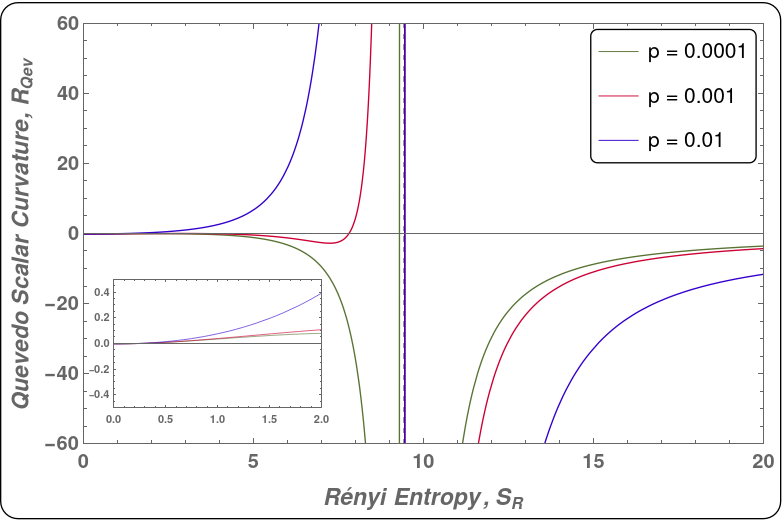}
	\caption{{\it\footnotesize Variations of Quevedo scalar curvature with R\'enyi entropy for different R\'enyi pressure. The vertical dashed line indicates a singularity at the entropy $S_0=3\pi Q^2 $. Electric charge is fixed at $Q=1$.}}
	\label{fig:fig6q}
\end{figure}

In the geometrothermodynamics framework, introduced by H. Quevedo \cite{Quevedo:2006xk,Quevedo:2007mj,Ladino2025geo}, thermodynamic systems are described using, unlike Ruppeiner metric, a Legendre-invariant metric structure on a phase  $\mathbb{T}$ that combines additive/extensive and intensive variables with the thermodynamic potential. In such a geometry, the line element adopting the Rényi entropy background is defined as
\begin{equation}
    ds^2=\Omega\left(\frac{\partial^2 M}{\partial Q^2}dQ^2-\frac{\partial^2 M}{\partial S_R^2}dS_R^2\right),
\end{equation}
where $\Omega$ is a conformal factor given by
\begin{equation}
    \Omega=S_R \frac{\partial M}{\partial S_R}+Q \frac{\partial M}{\partial Q}.
\end{equation}
One apparent advantage of the Quevedo metric over the Ruppeiner one is its diagonal structure. In Fig.\ref{fig:fig6q}, we plot the scalar curvature in the Quevedo geometry. Depending on the value of the pressure $p$, the scalar curvature changes sign, once for high pressure or twice for low pressure. As in the Ruppeiner case, this reveals the interactions between the black hole degrees of freedom take a repulsive nature at low R\'enyi entropy/small horizon radius and high pressure to an attractive nature as entropy/horizon radius increases to then swinging acutely in the neighborhood of $S_0=3 \pi Q^2$, from attractive to repulsive for high entropies/large horizon radius. In contrast, at lower pressures, the interactions start attractive at low entropies, then become highly repulsive around $S_0$ to regain a repulsive nature for higher values of entropy.

\begin{figure}[!ht]
 \vspace{0cm}
		\centering
		\begin{tabbing}
		\hspace{-2.1cm}
		\includegraphics[scale=0.51]{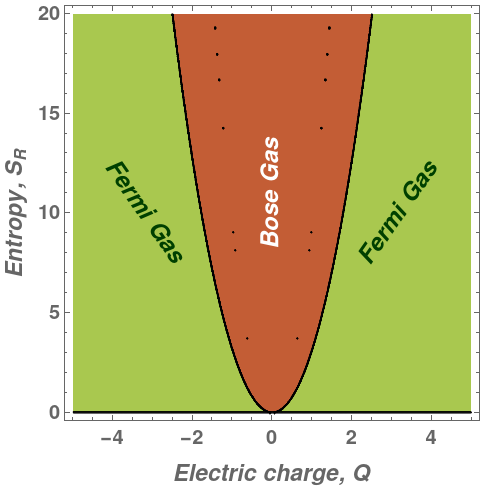}
		\hspace{-0.1cm}
		\includegraphics[scale=0.51]{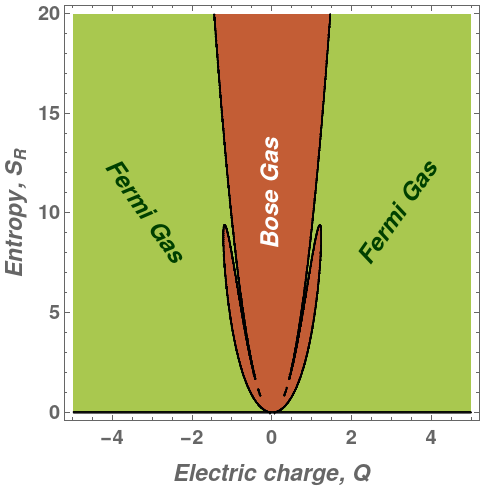}
            \hspace{-0.1cm}
		\includegraphics[scale=0.51]{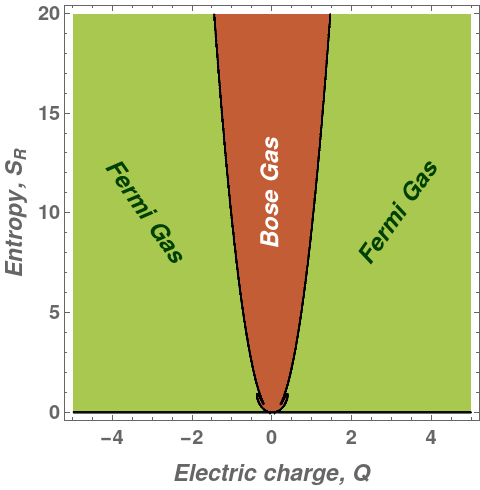}
		\end{tabbing}
		\vspace{-0.9cm}
		\centering
		\begin{tabbing}
		\hspace{-2.2cm}
		\includegraphics[scale=0.53]{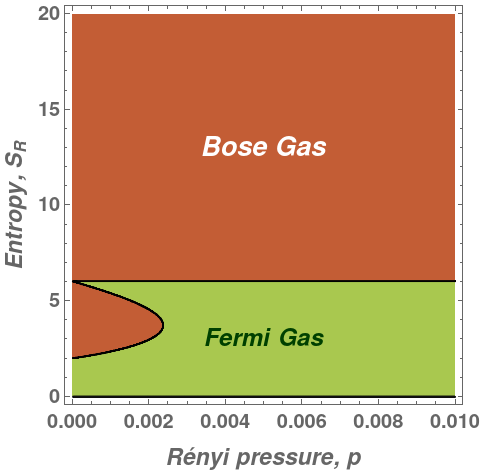}
		\hspace{-0.3cm}
		\includegraphics[scale=0.53]{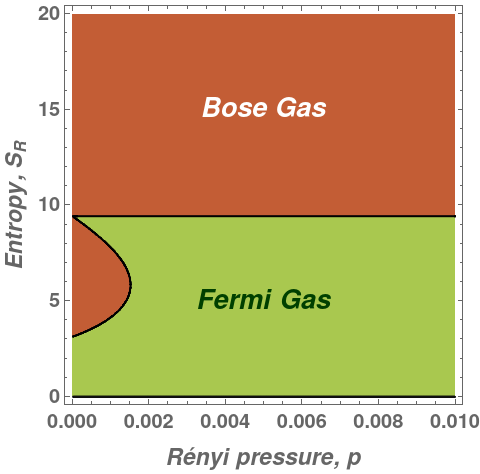}
            \hspace{-0.3cm}
		\includegraphics[scale=0.53]{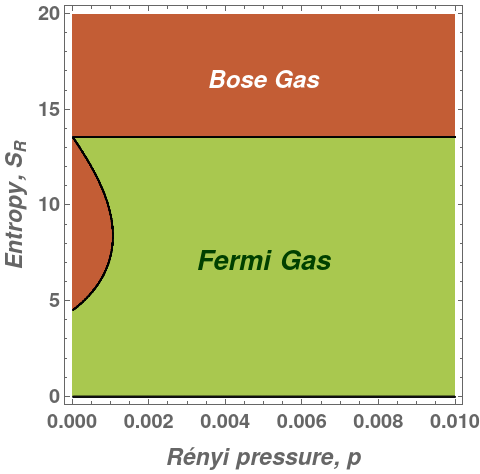}
		\end{tabbing}
		\vspace{-0.6cm}
		\caption{{\it\footnotesize \textit{\textbf{Top panel:}} Quevedo scalar curvature as a function of the R\'enyi entropy $S_R$ and the electric charge $Q$. \textbf{The orange region }indicates a bosonic gaz behavior $R_{Qev}<0$, \textbf{the green region }shows a fermionic gaz behavior $R_{Qev}>0$ and \textbf{the black line} denotes an ideal gaz comportment $R_{Qev}=0$. R\'enyi Pressure is increasing from left to right: $p=0.00001,\,0.001,\,0.01$. \textit{\textbf{Bottom panel:}} Quevedo scalar curvature as a function of the R\'enyi entropy $S_R$ and the R\'enyi pressure $p$. \textbf{The orange region }indicates a bosonic gaz behavior $R_{Qev}<0$, \textbf{the green region }shows a fermionic gaz behavior $R_{Qev}>0$ and \textbf{the black  line} denotes an ideal gaz comportment $R_{Qev}=0$. Electric charge is increasing from left to right: $Q=0.8,\,1,\,1.2$.}}
		\label{fig:density_quev}
	\end{figure}

A similar qualitative interpretation of the evolution of the Quevedo scalar curvature can be given: an ideal gas-like behavior at large entropies, an indication of a phase transition when the scalar curvature changes sign, a true singularity at $S_0=3\pi Q^2$ which may point to a breakdown of geometrothermodynamics in complete analogy with general relativity and an enfeebling of the interactions in the Planckian regime. However, Fig.\ref{fig:density_quev} invokes opposite conclusions in regard to the bosonic behavior as increasing nonextensivity effects seem to reduce the bosonic nature of the black hole fluid. This disagreement between the two metrics may be attributed to the thermodynamic potential dependence of Ruppeiner geometrothermodynamics, which induces a change of its metric signature following a Legendre transformation\cite{Quevedo:2006xk}.
% \newline
% \textcolor{red}{\small comments, plusieurs figures manquent de la même procedure}
\paragraph{} In the following section, we proceed to the quantization of the microscopic degrees of freedom of the R\'enyi flat black hole.

\section{Quantization of Black Hole Area in Rényi Formalism}\label{sect3}

% [10.1142/S0217751X22500361], [Dao-Quan Sun et al 2020 Class. Quantum Grav. 37 015008]\\
In this section, we investigate the horizon area of an asymptotically flat Rényi black hole from a quantization perspective, based on the idea of area pixelation as discussed in \cite{Ruppeiner2008}. In this framework, the black hole horizon is viewed as composed of discrete \textit{pixels}, each occupying one unit of Planck area, $l_p^2$. A group of $\gamma$ such pixels corresponds to a single microscopic degree of freedom, referred to as a black hole \textit{molecule}, which interacts with others in a manner analogous to molecules in a Van der Waals fluid.

In the Rényi thermodynamic framework \cite{Barzi:2023mit, Promsiri:2020jga, Wei:2015iwa}, the effective number density of black hole molecules is given by the inverse of the specific volume:
\begin{equation}\label{eq37}
\rho=\frac{1}{v}=\frac{N}{V}=\displaystyle \frac{3}{8 r_{h}l_p^2}.
\end{equation}
where the thermodynamic volume $V$ and $N$, the number of black hole "molecules" are defined as,
\begin{equation}\label{eq38}
V=\frac{4\pi r_h^3}{3}\quad \text{and} \quad 
N=\displaystyle \frac{4 \pi r_{h}^{2}}{\gamma}.
\end{equation}
Substituting into Eq.~\eqref{eq37}, we find that the number of pixels per molecule is $\gamma=8$.   
Furthermore, combining the expressions reveals that the number of molecules scales with volume as:
\begin{eqnarray}
   N =\frac{3}{8}\left(\frac{4\pi}{3}\right)^{\frac{1}{3}}V^{\frac{2}{3}} \propto V^{\frac{2}{3}}
\end{eqnarray}
This $2/3$ power-law dependence reflects the fundamentally surface-based nature of the black hole degrees of freedom. Consequently, any variation in the volume inherently affects the number of molecules, in contrast to ordinary thermodynamic systems where volume and particle number can be varied independently.

To explore area quantization, we assume the black hole horizon area is discretized as:
\begin{equation} 
A_n=\displaystyle  n\,\alpha\; l_p^2
\end{equation}
where $\alpha$ is the number of pixels constituting one quantum of area, measured in Planck units (we set $l_p^2=1$ hereafter for simplicity).

Under this quantization scheme, the ratio of the total number of area quanta to the number of molecules is:
\begin{eqnarray}\label{eq39s}
   \frac{\text{number of quanta of area}}{\text{number of molecules}}= \frac{\displaystyle\left(\frac{A_n}{\alpha}\right)}{ N}=\frac{A_n}{\alpha}\frac{\gamma}{A_n}=\frac{\gamma}{\alpha}\in\mathbb{N}
\end{eqnarray}
 which implies that $\gamma$ must be a multiple of $\alpha$, $\alpha|\gamma$\footnote{The symbol $|$ is read "\textit{divides}".}, knowing the value of $\gamma=8$, this condition yields the possible values:
\begin{equation}\label{eq44}
    \alpha = 1, 2, 4, \text{or }  8.
\end{equation}
These results suggest that black hole entropy and area may emerge from a quantized microscopic structure, with the ratio $\gamma/\alpha$ controlling how discrete "molecules" populate the horizon. The compatibility between Rényi thermodynamics and area quantization supports a statistical interpretation of the black hole degrees of freedom and reinforces the analogy with molecular systems in conventional thermodynamics.

%----------------------------

\begin{figure}[!ht]
 \vspace{0cm}
		\centering
		\begin{tabbing}
		\hspace{-2cm}
		\includegraphics[scale=0.46]{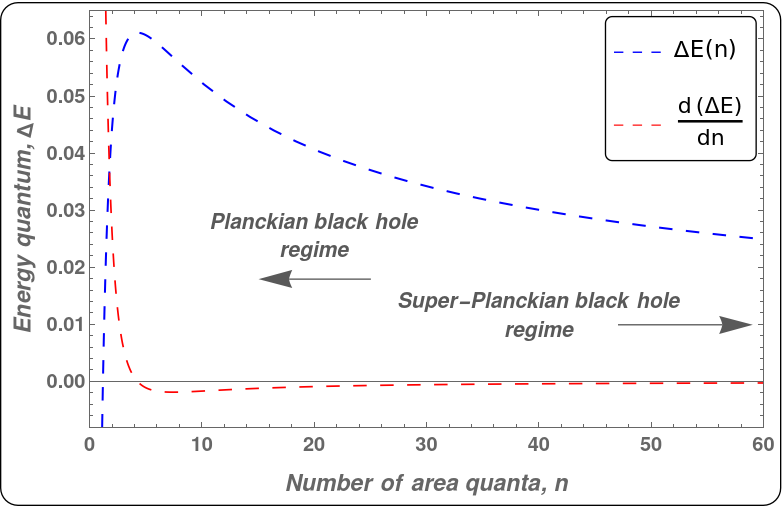}
		\hspace{0.1cm}
		\includegraphics[scale=0.47]{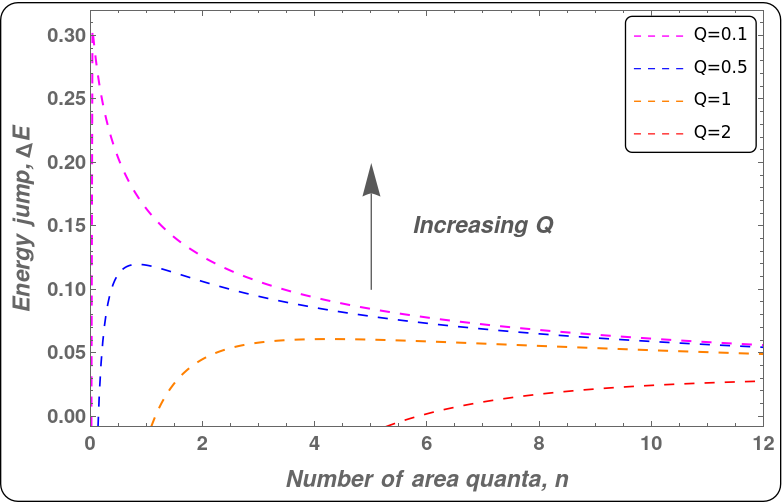}
		\end{tabbing}
		\vspace{-0.6cm}
		\caption{{\it\footnotesize Variations of energy quantum with the number of quanta of area. The left panel shows a maximum of positive energy quantum where the derivative (red dashed line) vanishes at $n\simeq4$ for $Q=1$. The right panel traces the evolution of the energy quantum with increasing electric charge. }}
		\label{fig:fig1}
	\end{figure}

The black hole mass, $\displaystyle M = \frac{r_h}{2}\left( 1 + \frac{Q^2}{r_h^2}\right)$, and horizon radius are also quantized such as,
% \begin{equation} 
% M=\displaystyle \frac{\frac{A}{4} + \pi Q^{2}}{\sqrt{\pi} \sqrt{A}}
% \end{equation}
\begin{equation}\label{eq42}
M_n=\displaystyle \frac{\alpha  n+4 \pi  Q^2}{4 \sqrt{\pi } \sqrt{\alpha  n}} \quad \text{and} \quad r_n=\sqrt{\frac{\alpha \,n}{4\pi}}
\end{equation}
From Eq.~\eqref{eq38}, we can derive a fundamental relation for the number of black hole molecules by substituting the quantized horizon radius $r_n$, which corresponds to the quantized area level $A_n=n\alpha l_p^2$. This yields
\begin{eqnarray}
    N\equiv N_n=\frac{\alpha}{\gamma}n
\end{eqnarray}
This relation reveals two key insights. First, the number of black hole molecules $N$ is directly proportional to the number of area quanta $n$. Second, for $N_n$ to be an integer for all $n$, the ratio $ \alpha/\gamma$ must also be an integer. This implies that $\gamma$ must divide $\alpha$, i.e., $\gamma|\alpha$. Given the allowed values from Eqs.~\eqref{eq39s} and \eqref{eq44}, the only consistent solution is
$\alpha = \gamma = 8$.  
As a result, the number of black hole molecules is exactly equal to the number of area quanta:
\begin{eqnarray}
    N=n.
\end{eqnarray}
Furthermore, the exchange of energy between the black hole and its surroundings is also quantized. The discrete energy jump associated with the absorption of a single area quantum (or, equivalently, the addition of one molecule) can be calculated as:
\begin{align} 
\Delta E&=M_{n+1}-M_n\\
&=\displaystyle \frac{\sqrt{n} \left[\alpha (n+1)+4 \pi  Q^2\right]-\sqrt{n+1} \left[\alpha  n+4 \pi  Q^2\right]}{4 \sqrt{\pi } \sqrt{\alpha  n (n+1)}}
\end{align}
Fig.\ref{fig:fig1} illustrates the variation of the energy quantum $\Delta E$ as a function of the number of area quanta $n$, along with its first derivative. These plots provide insight into the discrete thermodynamic behavior of the black hole at the microscopic level.
% \begin{figure}[!ht]
% 	\centering
% 	\includegraphics[scale=.7]{plot2_2.png}
% 	\caption{\footnotesize\it }
% 	\label{fig:fig1}
% \end{figure}
The energy quantum is positive for most of the values of $n$ which indicates the endothermic nature of the transitions $M_n\longrightarrow M_{n+1}$. It also exhibits a maximum at small $n\simeq4$ and an asymptote at very large $n\rightarrow\infty$ of the form
\begin{eqnarray}
    \Delta E =\displaystyle\frac{1}{\sqrt{ 8\pi n}}+O\left[\left(\frac{1}{n}\right)^{3/2}\right],
\end{eqnarray} 

Here, we have used $\alpha = 8$. This asymptotic behavior, which dominates in the regime of large $n$ (i.e., a large number of area quanta), is notably independent of the black hole’s electric charge $Q$. Furthermore, the energy quantum scales as $n^{-1/2}$, implying that as the black hole grows larger, \textit{in the super-Planckian regime}, it exchanges increasingly smaller energy quanta. In the large $ n$ limit, this leads to a quasi-continuous energy spectrum, reflecting the smooth classical behavior that emerges from the underlying discrete quantum structure. Besides, \textit{in the Planckian black hole regime}, the energy quantum increases to attain its maximum, which reveals the relative difficulty for small black holes to grow compared with larger ones. A notable feature is the negative value of the energy quantum at $n=1$ (for $Q=1$), meaning that the process $M_1\longrightarrow M_2$ is exothermic with may hint to a splitting of the quantum of area into two, releasing excess energy in the process. In the right panel of Fig.\ref{fig:fig1}, we depict the dependence of the energy quantum on the electric charge $Q$. As can be seen, the maximum of $\Delta E$ increases with decreasing charge; however, the endothermic region widens as the charge becomes larger.

% \begin{equation} 
% \begin{split}
%     &\frac{d\Delta E_{n\rightarrow n+1}}{dn}=\\\\ &\frac{4 \pi Q^2  \left[-\sqrt{n^3 (n+1)}+n (n+2)+1\right] 2+\alpha  \left[\sqrt{n^5 (n+1)}+\sqrt{n^3 (n+1)}-n (n+1)^2\right]}{8 \sqrt{\pi } \sqrt{\alpha } n^{3/2} (n+1)^2}
% \end{split}
% \end{equation}

% The maximum of energy quantum $\Delta E_{n\rightarrow n+1}$ occurs at,

% \begin{equation} 
% n_{max}=\displaystyle \bigg\lfloor\frac{192 \pi  Q^2-17 \alpha+\sqrt{3} \sqrt{75 \alpha ^2+12288 \pi ^2 Q^4+640 \pi  \alpha  Q^2}}{32 \alpha }\bigg\rfloor
% \end{equation}
% Where $\lfloor x\rfloor$ is the integer part of $x$, and takes the value,
% \begin{equation} 
% \left(\Delta E_{n\rightarrow n+1}\right)_{max}= -\frac{
% \begin{split}
%  24576 \pi ^2 Q^4+64 \pi  Q^2 \left(51 \alpha +2 \sqrt{3} a\right)+\alpha  \left(7 \sqrt{3} a-135 \alpha \right) 
% \end{split}
% }{16 \sqrt{2 \pi } \left(15 \alpha +128 \pi  Q^2\right)}
% \end{equation}
% Where $a=\sqrt{75 \alpha ^2+12288 \pi ^2 Q^4+640 \pi  \alpha  Q^2}$.
The number density for a state with $n$ quanta of area (or $N$ molecule) is given by the expression of $r_n$, Eq.\eqref{eq42} in Eq.\eqref{eq37},
\begin{equation}\label{eq_rho_n}
\rho_n=\displaystyle \frac{6 \sqrt{\pi}}{\gamma \sqrt{\alpha} \sqrt{n}}
\end{equation}

At the critical equilibrium state, corresponding to the critical temperature $T = T_c$ \cite{Barzi:2023mit}, a second-order phase transition occurs between the \textit{small black hole (SBH)} and the \textit{large black hole (LBH)} phases. Above this temperature, the distinction between the two phases disappears, and the system enters a single, unified phase. For the charged-flat Rényi black hole, the critical horizon radius associated with this transition is given by
\begin{equation} 
r_c=\sqrt{6}Q,
\end{equation}
Which leads to a critical effective number density
\begin{equation} 
\rho_c=\displaystyle \frac{3}{8 \sqrt{6} Q},
\end{equation}
and to the critical number of quanta, taking $\gamma=8$ is obtained to be
\begin{equation}
    n_c=N_c=\frac{4\pi r_c^2}{\gamma l_p^2}=\frac{3\pi Q^2}{l_p^2},
\end{equation}
where the Planck area is made explicit for clarity. 
% \begin{equation} 
% n_c=\displaystyle \frac{36 \pi}{\gamma^{2} \alpha \rho_{c}^{2}}=\displaystyle \frac{1536 \pi Q^{2}}{\gamma^{2} \alpha}
% \end{equation}

% To reproduce the same critical number of quanta  manifested by the charged AdS black hole, we take $\alpha=4$ and $\gamma=8$,

% \begin{equation} 
% n_c=\displaystyle 6 \pi Q^{2}
% \end{equation}

% which is the same value as for GB charged AdS black hole independent of statistics.
\subsection{R\'enyi number of microstates}
\paragraph{}Given that $N, (=n)$ represents the number of microscopic degrees of freedom (\textit{dof}) of the black hole in a macrostate consisting of $N$ molecules, and assuming all \textit{dof} are equivalent due to the system's perfect symmetry, the number of compatible microstates, $\Omega_n$, must satisfy the inequality
\begin{equation}\label{eq53}
\Omega_n\leq\theta^{N},
\end{equation}
where $\theta$ denotes the average number of individual microscopic configurations accessible to each \textit{dof}. From the Rényi formalism, the entropy can be related to the number of microstates by $S_R = \ln \Omega_n$, as given in Eq.\eqref{eq9}, yielding
\begin{align}\label{eq55}
\displaystyle\Omega_n&=\left(1+\lambda S_{BH}\right)^{\displaystyle\frac{1}{\lambda}}=\left(1+ \frac{\lambda}{4}A_n\right)^{\displaystyle\frac{1}{\lambda}},\\
&=\left(1+ \frac{\lambda}{4}\alpha \,n\right)^{\displaystyle\frac{1}{\lambda}}.\label{ea55}
\end{align}
Equating Eqs.\eqref{eq53} and \eqref{eq55}, we find the condition
\begin{equation} 
\theta\geq\left(1+ \frac{\lambda}{4}\alpha \,n \right)^{\displaystyle \frac{\gamma}{\alpha\,\lambda\, n}}.
\end{equation}
Expanding for small $\lambda \ll 1$ gives the approximation:
\begin{equation} 
\theta\geq\displaystyle \left(1 - \frac{\gamma \alpha \lambda n l_p^2 }{32}\right)\exp\left({\frac{\gamma}{4}}\right) + O\left(\lambda^{2}\right).
\end{equation} 
Substituting $\alpha = \gamma = 8$, we obtain
\begin{equation}\label{theta}
\theta\geq\displaystyle e^2\left(1 -  2\lambda\, n \right). 
\end{equation}
Eq.\eqref{theta} reveals that in the Boltzmann-Gibbs (extensive) limit, $\lambda \to 0$, the minimal average number of microstates per \textit{dof} is $(\theta_{GB}){\text{min}} = e^2 \simeq 7$, regardless of the black hole's state. However, within Rényi statistics ($\lambda > 0$), nonextensive effects reduce the number of accessible microstates as $n$ increases. Beyond a certain threshold $n_{\text{lim}}$, the system can no longer accommodate additional microstates
\begin{equation} 
n_{lim}=\frac{1}{2\lambda \,l_p^2},
\end{equation}
This corresponds to a maximal horizon area
\begin{equation} 
A_{lim}=\frac{\alpha}{2\lambda}=\frac{4}{\lambda}.
\end{equation}
Unlike in extensive thermodynamics, where a black hole can grow indefinitely, the presence of nonextensivity introduces a universal upper bound on black hole size,
\begin{eqnarray}
    r_{lim}=\frac{1}{\sqrt{\pi\lambda}}
\end{eqnarray}
From an observational standpoint, we expect supermassive astrophysical black holes \cite{patrick2017ob,Shemmer2004ob,Kozłowski2017ob} to reach horizon radii close to this theoretical limit. For a rotating black hole with spin parameter $a$, the maximal observed horizon radius is
\begin{eqnarray}
    r_{lim}^{obs}=\sqrt{\frac{1}{\pi\lambda}-a^2}\approx r_{lim}.
\end{eqnarray}
This observational constraint provides a lower bound on the nonextensivity parameter $\lambda$:
\begin{eqnarray}
  \lambda\geq\lambda_{lim}=\frac{1}{\pi\,(r_{lim}^{obs})^2}.
\end{eqnarray}
Astronomical observations\cite{Ge2019ob,Shemmer2004ob} have positioned the supermassive black hole at the center of the quasar \textit{TON 618} as the largest black hole ever recorded so far, with a mass $M$, $\displaystyle\log\left(\frac{M}{M_\odot}\right)=10.82$. This puts its horizon radius at approximately
\begin{eqnarray}
    r_{lim}^{obs}=\displaystyle 2.24 \times 10^{67.82}\;GeV.
    %\frac{4\, 10^{40.82} }{1.7827 \,10^{-27}}
\end{eqnarray}
Which gives an estimate for the lower bound on nonextensivity parameter $\lambda$ as
\begin{eqnarray}
   \lambda_{lim}=9.54 \times10^{-70} \;GeV^{-2}
\end{eqnarray}
In other words, this lower bound on nonextensivity determines how large black holes can grow: the smaller the $\lambda$, the larger the maximal black hole size. 
\paragraph{}Assuming the conjectured duality between four-dimensional asymptotically AdS black holes in Boltzmann-Gibbs statistics and asymptotically flat black holes in Rényi thermodynamics \cite{Czinner:2015eyk,Barzi:2023mit,Barzi2024yan,Barzi2024yin}, we can also estimate an upper bound on the cosmological constant of AdS spacetime ($\Lambda < 0$) as
%In other words, this lower bound on nonextensivity establishes a limit on how large black holes might grow such that the smaller the bound on $\lambda$ the larger the limit on black hole size. Recalling the conjectured equivalence between the four-dimensional asymptotically AdS black hole in Boltzmann-Gibbs statistics and the asymptotically flat black hole with R\'enyi entropy\cite{Czinner:2015eyk,Barzi:2023mit,Barzi2024yan,Barzi2024yin}, we compute also an upper bound on the gravitational constant of AdS space $\Lambda<0$ as,
\begin{eqnarray}
    \Lambda_{lim}=-\frac{3\pi}{4}\lambda_{lim}=-2.25\times10^{-69} \;GeV^{-2}.
\end{eqnarray}
%-------------------------------
%To give an intuitive picture for this value, $\lambda_{lim}$ is approximately the number of Planck's areas in a

\paragraph{}The number of microstates of the black hole can be written as $\Omega_n=2^{B_n}$, where $B_n$ is the number of bits in the sense of Shannon's theory of information. These bits of information serve to encode the microstates of the black hole. In fact, what we call entropy is nothing more than this number of bits $B\propto S$. Given the quantized formula of the R\'enyi entropy,
\begin{eqnarray}
    S_n=\frac{1}{\lambda}\log\left(1+\frac{\lambda\alpha\,n}{4}\right).
\end{eqnarray}
We deduce the number of bits of information in the R\'enyi charged-flat black hole as ($\alpha=8$)
\begin{eqnarray}\label{eq_72bit}
    B_n=\frac{1}{\lambda}\log_2\left(1+2\lambda\,n\right).
\end{eqnarray}
For small nonextensivity parameter $\lambda$,
\begin{equation}
    B_n=\frac{2n}{\log(2)}-\frac{\lambda n^2}{\log(2)}+O\left(\lambda ^2\right).
\end{equation}
This expression reveals that in the extensive Boltzmann–Gibbs limit ($\lambda \rightarrow 0$), the number of bits is   $\displaystyle \frac{2}{\log(2)}\simeq 6.6439$ times the number of area quanta $n$ (or equivalently, the number of molecules $N$). However, in the nonextensive Rényi case ($\lambda > 0$), $B_n$ decreases as $n$ increases, indicating a suppression of independent degrees of freedom due to statistical correlations. As $n$ approaches the critical value $\displaystyle n \simeq \frac{2}{\lambda\log(2)}$, the number of bits $B_n$ tends toward zero, implying that all degrees of freedom become fully correlated, and the system effectively collapses into a single microstate: $B_n \to 0$, $\Omega_n \to 1$. This collective and perfectly coherent behavior is reminiscent of quantum phenomena such as the Bose–Einstein condensate or the superfluid phase of helium-4. It is important to note that the number of bits computed above is an exponent that gives the number of microstates of the black hole. However, to encode these microstates, one needs an integer number of bits\cite{Barzi2025info}, that is, the \textit{actual} number of bits to encode \textit{uniquely} all the microstates of the black hole is $\displaystyle\bigg\lceil\frac{2n}{\log(2)}\bigg\rceil$, where $\big\lceil.\big\rceil$ is the ceiling function.

Notably, the factor of $2$ in $\displaystyle \frac{2}{\log(2)}$ bits per black hole molecule arises from the ratio $\displaystyle\frac{\alpha}{4}$, where $\alpha = 8$ is the number of Planck area units per molecule. This implies that $\displaystyle8n$ Planck areas accommodate $\displaystyle\bigg\lceil\frac{2n}{\log(2)}\bigg\rceil$ bits of information or equivalently,$\displaystyle\bigg\lceil\frac{2n}{\log(2)}\bigg\rceil$ share $8n$  Planck areas. Thus, each bit of information occupies on average a number of Planck areas given by the \textit{Planck-per-Bit measure} denoted by $PpB$. In Gibbs-Boltzmann (GB) statistics it reads,
\begin{equation}
    PpB_{(GB)}=\displaystyle\frac{8n}{\bigg\lceil\displaystyle\frac{2n}{\log(2)}\bigg\rceil}\overset{n\rightarrow\infty}{\longrightarrow}4\log(2)\simeq2.77259
\end{equation}
In R\'enyi (R) statistics, the occupancy is generalized to,
\begin{equation}
   PpB_{(R)}=\displaystyle\frac{8n}{\bigg\lceil\displaystyle\frac{\log_2\left(1+2\lambda\,n\right)}{\lambda}\bigg\rceil}.
\end{equation}
\begin{figure}[!htb]
	\centering
	\includegraphics[scale=0.65]{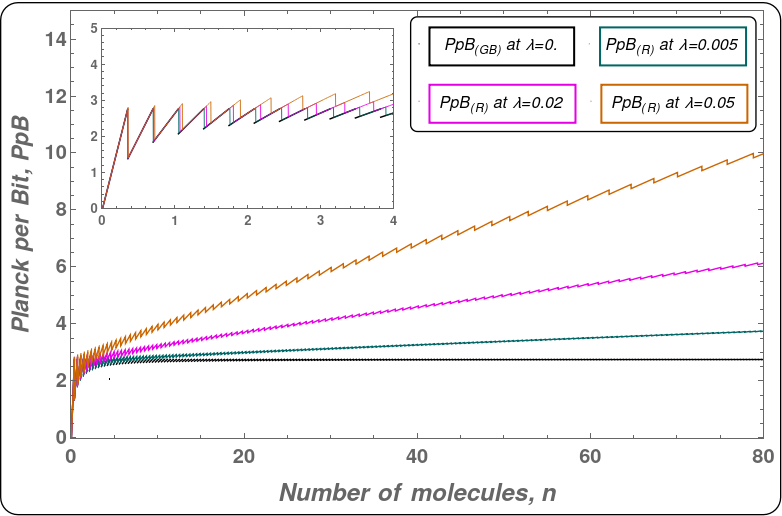}
	\caption{{\it\footnotesize Planck-per-Bit (PpB) of the bits of information as a function of the number of molecules of the Gibbs Boltzmaniann and R\'enyian statistics. The black line indicates the Boltzmaniann $PpB$, while colored lines depict R\'enyian $PpB$. The inset plot shows the discontinuity in the $PpB$. }}
	\label{fig:fig_PpB}
\end{figure}
And for small $\lambda\ll1$ it becomes,
\begin{equation}
   PpB_{(R)}=\displaystyle\frac{8n}{\bigg\lceil\displaystyle\frac{2n}{\log(2)}-\frac{\lambda n^2}{\log(2)}\bigg\rceil}.
\end{equation}

In Fig.\ref{fig:fig_PpB}, we depicted the variations of the Planck-per-Bit ($PpB$) quantity within the two statistics. It is clear that $PpB_{GB}$ saturates while $PpB_{R}$ increases monotonically with the number of black hole molecules. This shows that nonextensivity permits bits of information to be encoded on an increasing number of Plank areas. Thus reducing the informational content of each Planck area/pixel. $PpB^{-1}$ measures such content. A remarkable feature is the \textit{spiking} displayed by the $PpB$ as the number of bits necessary to encode the microstates rises with the number of molecules. 

In the limit $\lambda \ll 1$, we can invert the Rényi entropy expression to obtain a quantized relation for the black hole horizon area in terms of the number of information bits, Eq.\eqref{eq_72bit}:
\begin{align}
  \displaystyle  &\frac{A_n}{4} =\displaystyle \frac{2^{\lambda B_n}-1}{\lambda},\\
   & A_n=\displaystyle  4\frac{d}{d\lambda}2^{\lambda B_n}\Bigg|_{\lambda=0},\\
    &\simeq 4B_n l_p^2 \:(\lambda\rightarrow 0).\label{eq_78a}
\end{align}
Equation \eqref{eq_78a} thus recovers, in the limit of vanishing nonextensivity, the standard Bekenstein–Hawking relation between the number of bits of \textit{information stored in the black hole} and the horizon area in its quantized form.

%\textcolor{red}{\huge Here}
\subsection{Phase transition in the quantized R\'enyi formalism}
One can investigate the effect of a change in the number of area quanta, $n \rightarrow n+1$, on the entropy, or equivalently, on the number of bits of information $B_n$. The change in entropy is given by
\begin{eqnarray}
    \Delta S_n=\frac{1}{\lambda}\log\left(\frac{4+(n+1)\lambda\alpha}{4+n\lambda\alpha}\right).
\end{eqnarray}
For small $\lambda\ll1$ we get,
\begin{eqnarray}
    \Delta S_n=\frac{\alpha }{4}-\frac{\alpha^2\lambda  }{32} \left(1+2 \displaystyle n\right)+O\left(\lambda ^2\right).
\end{eqnarray}
The Rényi charged-flat black hole exhibits a first-order Van der Waals–like phase transition between a liquid-like \textit{small black hole (SBH)} phase and a gas-like \textit{large black hole (LBH)} phase. During this transition, entropy (or the number of information bits) undergoes a discontinuous jump. In the quantized formalism, this entropy change is evaluated to be 
\begin{align}
   \Delta S_{L\rightarrow S}&=S_{SBH}-S_{LBH}\\
   &= \frac{\alpha}{32}  \left(n_S-n_L\right) \left[8-\alpha\lambda  \left(n_L+n_S\right)\right],\label{eq69}
\end{align}
where $n_L$ and $n_S$ are the numbers of quanta of area/molecules corresponding to the LBH and SBH phases, respectively. Then, the latent heat of this first-order phase transition is defined through the Clapeyron relation,
\begin{equation}\label{bhr22}
\frac{dP_0}{dT_0}=\frac{L_{L\longrightarrow S}}{T_0(v_S-v_L)},
\end{equation}
where $P_0$ and $T_0$ are the coexistence pressure and temperature. $v_L$ and $v_S$ are the specific volumes of the LBH and SBH states, respectively. In quantized form, the latent heat becomes
\begin{align}\label{eq66}
L_{L\rightarrow S}
&=\displaystyle \frac{T_{0} \gamma \sqrt{\alpha} \left(\sqrt{n_{S}} - \sqrt{n_{L}}\right)}{6 \sqrt{\pi}}\;\frac{dP_0}{dT_0}.
\end{align}
From Eq.~\eqref{eq66}, we see that the latent heat originates from the discontinuity in the number of area quanta $n$ between the two phases. Since $n_L > n_S$, a rearrangement or coalescence of area quanta (or molecules) occurs during the $LBH \rightarrow SBH$ transition. 
Alternatively, the latent heat can be defined directly from the entropy change:
\begin{eqnarray}\label{eq72}
    L_{L\rightarrow S}=\frac{T_0 \Delta S_{L\rightarrow S}}{N_{av}}
\end{eqnarray}
where $N_{av}$ is the average number of black hole molecules. This average may be taken as a simple arithmetic mean, $\displaystyle N_{av} = \frac{N_L + N_S}{2}$, or a more elaborate functional form. In all cases, it must satisfy $N_S\leq N_{av}\leq N_L$, thus, the latent heat must also be bounded
\begin{eqnarray}
   \frac{T_0 \Delta S_{L\rightarrow S}}{N_{L}} \leq L_{L\rightarrow S}\leq\frac{T_0 \Delta S_{L\rightarrow S}}{N_{S}}.\label{eq72a}
\end{eqnarray}
Using Eqs.\eqref{eq72} and \eqref{eq69}, the latent heat is expressed as
\begin{eqnarray}
     L_{L\rightarrow S}=\frac{\alpha T_0\left(n_S-n_L\right)}{32N_{av}} \left[8-\alpha \lambda \left(n_L+n_S\right)\right].
\end{eqnarray}
Replacing $\alpha=8$ and $N_{av}=n_{av}$, we get
\begin{eqnarray}
     L_{L\rightarrow S}=\frac{2 T_0\left(n_S-n_L\right)}{n_{av}}    \left[1- \lambda \left(n_L+n_S\right)\right].\label{eq76}
\end{eqnarray}
One sees that the latent heat vanishes for $n_L=n_S$, which corresponds to the critical phase transition $n_L=n_S=n_c$, but also for an intriguing special case occurring when $n_{tot}=n_L+n_S$ reaches the threshold value of $\displaystyle\frac{1}{\lambda l_p^2}$. There, the phase transition $LBH\longrightarrow SBH$ becomes an {\it athermal} process, $L_{L\rightarrow S}=0$, then endothermic, $L_{L\rightarrow S}<0$, if $n_{tot}$ exceeds this threshold. From \cite{Barzi:2023mit} we have the expressions of coexistence temperature, $T_0$, and latent heat in terms of the ratio,  $\displaystyle\omega=\frac{r_S}{r_L}=\sqrt{\frac{n_S}{n_L}}$,
\begin{equation}\label{eq77a}
T_0(\omega)= \frac{\omega \left(\omega + 1\right)}{\pi Q \left(\omega^{2} + 4 \omega + 1\right)^{\frac{3}{2}}} \quad \text{and} \quad L(\omega)= \frac{3\omega (1-\omega )   (\omega +1)^2 }{2 \pi  Q\left(\omega ^2+\omega +1\right) (\omega^2+4\omega+1)^{3/2}}.
\end{equation}
The parameter $0\leq\omega\leq1$ completely parametrizes the phase transition $LBH\longleftrightarrow SBH$, with $\omega=1$ representing the critical second-order phase transition.   
\paragraph{}With the help of these expressions, one can find a formula for the average $n_{av}$. By injecting Eqs.\eqref{eq77a} and \eqref{eq76} and solving for $n_{av}$,
\begin{align}
   n_{av}&=N_{av}=\frac{4 \left(\omega ^2+\omega +1\right) }{3 \left(1-\omega ^2\right)}\left(n_L-n_S\right) \left[1-\lambda  \left(n_L+n_S\right)\right],\\
   &=\frac{4 \left(\omega ^2+\omega +1\right) }{3 (\omega +1)} n_L\left[1-\lambda  (\omega +1) n_L\right].\label{eq83}
\end{align}
Therefore, in Eq.\eqref{eq72}, the $N_{av}=n_{av}$ that reproduces the Clapeyron relation for the coexistence curve, Eq.\eqref{eq66}, is not a simple arithmetic or geometric averaging, but a complicated one as given by Eq.\eqref{eq83}. Furthermore, this average vanishes for some nontrivial values of the $n_L$ and $n_S$,
\begin{equation}
   n_L=\frac{1}{\lambda  (\omega +1)}\quad \text{and}\quad n_S=\frac{\omega}{\lambda  (\omega +1)}.
\end{equation}
However, the entropy change $\Delta S_{L\rightarrow S}$, Eq.\eqref{eq69} also vanishes for the same values. 
\subsection{Partition Function in The Quantized Formalism}
\paragraph{}In Boltzmaniann statistical mechanics, the partition function $Z_{BG}^{(q)}$ of the quantized Schwarzschild flat black hole in the canonical ensemble is given by,
\begin{equation}\label{eq82a}
Z_{BG}^{(q)}=\sum_{n=0}^{\infty} \exp\left(- \beta M_n\right),
\end{equation}

where $M_n=\displaystyle \frac{\sqrt{\alpha  n}}{4 \sqrt{\pi } }$, is the quantized mass given by Eq.\eqref{eq42} for $Q=0$. This formulation of the partition function is consistent with the standard Boltzmann–Gibbs formalism and does not incorporate any nonextensive corrections. 
In contrast, in \cite{Barzi:2024vvo}, we computed the partition function of the continuous (unquantized) Schwarzschild Rényi-flat black hole using a generalized Euclidean path integral approach combined with Wick rotation. This formalism naturally incorporates the nonextensive nature of black hole thermodynamics within the Rényi framework, allowing for a more complete statistical description beyond the Boltzmannian limit
\begin{equation}\label{eq_21}
Z^{(nq)}_\lambda=\exp\left(\displaystyle -\frac{\beta^{2}}{16 \pi}-\frac{\lambda \beta^{4}}{512 \pi^{2}}\right),
\end{equation}
% Where $T_n$ is the R\'enyi temperature in the $n^{th}$ state,
% \begin{eqnarray}
%     T_n=\frac{(\alpha  \lambda  n+4  ) \left(\alpha  n-4 \pi  Q^2\right)}{8 \pi ^{1/2} (\alpha  n)^{3/2}}
% \end{eqnarray}
% The partition function is found to be,
% \begin{align}
%     Z=\displaystyle \sum_{n=0}^{\infty}\exp \left(\frac{2 \pi  \alpha  n \left(\alpha  n+4 \pi  Q^2\right)}{(\pi  \alpha  \lambda  n+4 \pi ) \left(4 \pi  Q^2-\alpha  n\right)}\right)
% \end{align}y
where we used the R\'enyi-Wick's rotation defined also in \cite{Barzi:2024vvo} as
\begin{equation}\label{eq_ren_wick}
t\rightarrow-i\left(\displaystyle \tau+ \frac{\lambda \tau^{3}}{48 \pi}\right).
\end{equation}
A close inspection of the partition function $Z_\lambda^{(nq)}$ shows that nonextensivity effects will dominate the statistical behavior of the Schwarzschild black hole at inverse temperatures larger than a characteristic value $\beta_\lambda$ given by,
\begin{equation}
    \beta_\lambda=4\pi\sqrt{\frac{2}{\lambda}}.
\end{equation}
For small $\lambda\ll1$, the corresponding temperature $1/\beta_\lambda$ is equally small. This indicates that nonextensivity plays a dominant role in super-Planckian black holes.

A natural generalization to nonextensive statistical mechanics is performed in the quantized case where Eq.\eqref{eq82a} is replaced, through the R\'enyi-Wick rotation Eq.\eqref{eq_ren_wick} by,
\begin{eqnarray}\label{eq_21q}
    Z^{(q)}_\lambda=\sum_{n=0}^{\infty} \exp\left(- \beta M_n-\frac{\lambda }{48 \pi} \beta^{3}M_n\right).
\end{eqnarray}
\begin{figure}[!ht]
	\centering
	\includegraphics[width=0.7\linewidth]{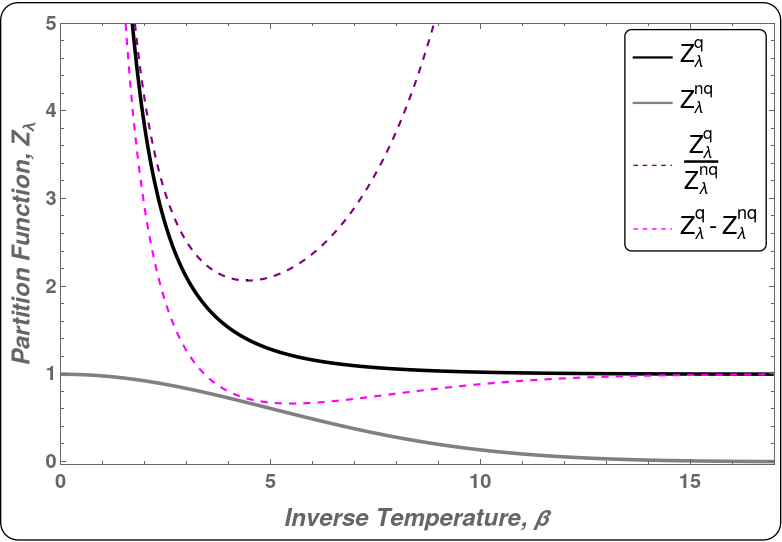}
	\caption{\it \footnotesize Partition function of the quantized $(q)$ (Blue) and unquantized $(nq)$ (Black) asymptotically flat Schwarzschild black hole within R\'enyi statistics. We fixed $\lambda=0.001$ and $\alpha=8$.}
	\label{fig:fnZ}
\end{figure}
The quantized probability distribution of the black hole energy states is calculated to be
\begin{align} 
 \mathcal{P}^\lambda_n&=\frac{\displaystyle\exp \left(-\beta M_n-\frac{\lambda }{48 \pi} \beta^{3}M_n\right)}{Z^{(q)}_\lambda},\\
 &= \displaystyle \left(Z^{(q)}_\lambda\right)^{-1}\exp \left[- \frac{ \beta\sqrt{n}}{\sqrt{2\pi }}-\frac{\lambda\beta^3\sqrt{n}}{48\sqrt{2\pi^3 }}\right],
\end{align}
here the value of $\alpha=8$ was used in $M_n$. It is worth noting that in the limit $\lambda\rightarrow0$, we recover the extensive Boltzmann-Gibbs probability distribution function for the asymptotically flat Schwarzschild black hole as
\begin{eqnarray}
   \mathcal{P}^0_n= \displaystyle \left(Z^{(q)}_0\right)^{-1}\exp \left[- \frac{ \beta\sqrt{n}}{\sqrt{2\pi }}+\frac{\beta^{2}}{16 \pi}\right].
\end{eqnarray}
In Fig.\ref{fig:fnZ} we plotted the quantized $Z^q_\lambda$ and the unquantized $Z^{nq}_\lambda$ partition functions of the R\'enyi Schwarzschild black hole. Both functions exhibit strikingly different behavior in low and high temperature limits. At high temperatures, $\beta\rightarrow0$, the unquantized partition function saturates at the value one as expected for classical thermodynamic systems in a heat bath that presents an infinite amount of thermal agitation, where no physical structure can persist and only the thermal radiation phase is expected to dominate. The quantized function, on the other hand, becomes infinite as all the black hole quantum states are equally probable and since there is an infinite number of them, their some are infinite. In the low temperature limit, $\beta\rightarrow\infty$, the two partition functions saturate at different values, zero for the unquantized form and at one for the quantized form. A vanishing value of the partition function traduces the very low probability of any of the states of the black hole to manifest itself, while a finite value indicates that the black holes phase may persist at very low temperatures.

\begin{figure}[!ht]
 \vspace{0cm}
		\centering
		\begin{tabbing}
		\hspace{-2.1cm}
		\includegraphics[scale=0.46]{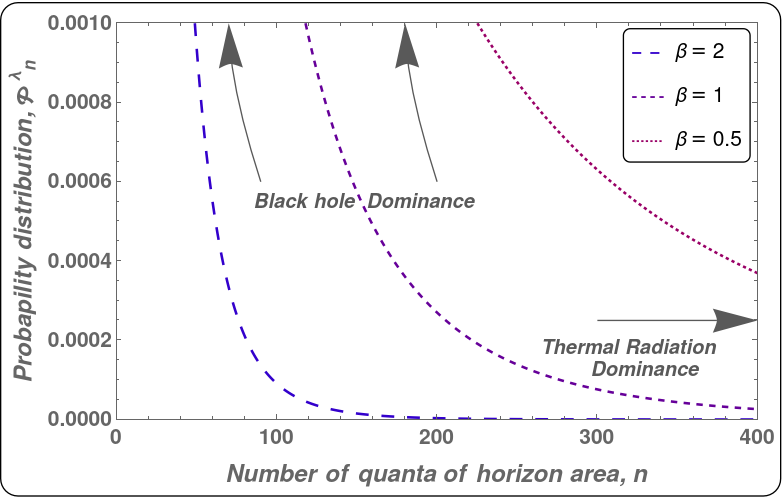}
		\hspace{0.1cm}
		\includegraphics[scale=0.48]{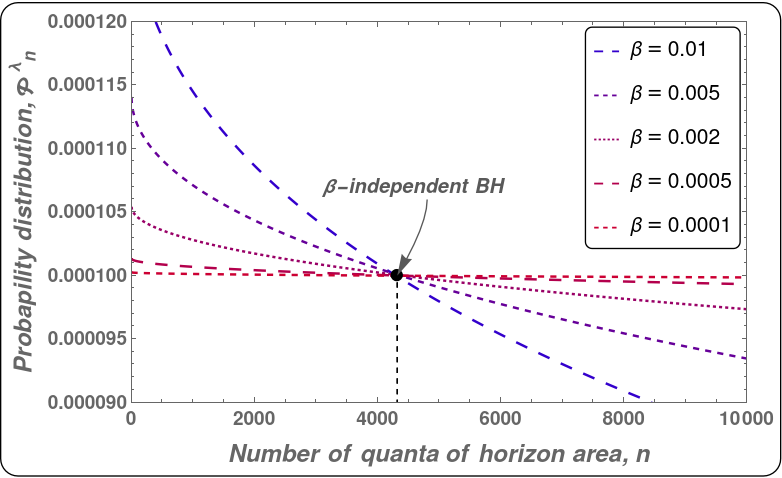}
		\end{tabbing}
		\vspace{-0.6cm}
		\caption{{\it\footnotesize Quantized probability distribution of Schwarzschild R\'enyi-flat black hole in the canonical ensemble. We fixed $\lambda=0.001$ and $\alpha=8$.}}
		\label{fig:fncc}
	\end{figure}

\paragraph{}We depicted in Fig.\ref{fig:fncc} the variations of the probability distribution of the quantized asymptotically flat Schwarzschild black hole in R\'enyi statistics as a function of the number of quanta of area $n$. It is noted that at low temperatures, the probability distribution function rapidly decreases with $n$. Moreover, a moderate increase in temperature profoundly transforms the probability distribution, which increases significantly as $\beta\geq0.5$. The probability distribution increases rapidly for small $n$, suggesting that small and Planckian black holes are much more probable than large ones. On the right panel of Fig.\ref{fig:fncc}, a new feature is revealed at high temperatures $\beta\leq0.01$, where a particular black hole state seems to have the same probability independently of the heat bath temperature. 

\begin{figure}[!ht]
	\centering
	\includegraphics[width=0.7\linewidth]{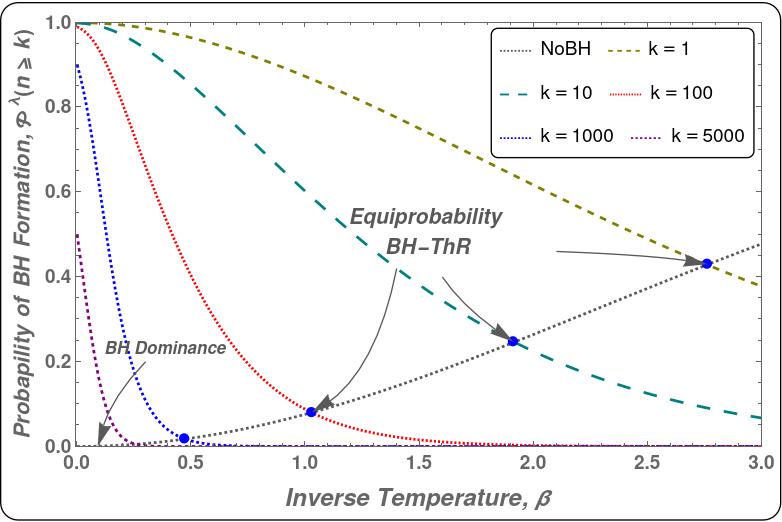}
	\caption{\it \footnotesize Probability of the formation of the quantized Schwarzschild R\'enyi-flat black hole in the canonical ensemble (cumulative distribution function). The values of $k$ represent the minimum number of black hole quanta of area. The gray dashed curve indicates the probability of non formation of the quantized black hole. We fixed $\lambda=0.001$ and $\alpha=8$.}
	\label{fig:fn_b}
\end{figure}

In Fig.\ref{fig:fn_b} we portrayed the cumulative distribution function $\mathcal{P}^\lambda(n\geq k)$, which measures the probability of the formation of a R\'enyi-Schwarzschild black hole of at list $k$ quanta of area. Plotted also is the probability that only thermal radiation (ThR) exists (NoBH curve). It is shown that the black hole phase dominates at high temperatures $\beta\rightarrow0$ while thermal radiation dominates at low temperatures, $\beta\rightarrow\infty$, in addition we indicated points of equal probability to having thermal radiation or a black hole phases.

\paragraph{}The average number of quanta $\langle n \rangle_\lambda$ is computed such as,
\begin{align}
\langle n \rangle_\lambda &=\sum_{n=0}^{\infty}n\,\mathcal{P}^\lambda_n,\\
&=\left(Z^{(q)}_\lambda\right)^{-1}\sum_{n=0}^{\infty}n \displaystyle \exp \left[- \frac{ \beta\sqrt{n}}{\sqrt{2\pi }}-\frac{\lambda\beta^3\sqrt{n}}{48\sqrt{2\pi^3 }}\right],
% &=-\left(\frac{Z_\lambda^{-1}\sqrt{\pi}}{2}\right)\frac{d}{d\beta}\left( \frac{\displaystyle 1 }{\displaystyle 1 - \exp\left(-\frac{2 \beta}{\sqrt{\pi }}\right)}\right)\\
\end{align}

% Recalling the energy of the Schwarzschild BH, 
% \begin{equation} 
% M_n=\displaystyle \frac{\sqrt{\alpha} \sqrt{n}}{4 \sqrt{\pi}}
% \end{equation}
while the average mass, $\langle M \rangle_\lambda$, is defined as,
\begin{eqnarray}
    \displaystyle \langle M\rangle_\lambda =\sum_{n=0}^{\infty}M_n\,\mathcal{P}^\lambda_n.
\end{eqnarray}
Which in turn gives the average mass of the quantized black hole as,

\begin{align}
\langle M \rangle_\lambda^{(q)}&=\displaystyle \frac{\left(Z^{(q)}_\lambda\right)^{-1}}{\sqrt{2\pi }}\sum_{n=0}^{\infty}  \sqrt{n}\displaystyle \exp \left[- \frac{ \beta\sqrt{n}}{\sqrt{2\pi }}-\frac{\lambda\beta^3\sqrt{n}}{48\sqrt{2\pi^3 }}\right].
\end{align}

On the one hand, the average mass of the unquantized black hole is defined from statistical mechanics, such as,
\begin{align}
\langle M \rangle_\lambda^{(nq)}&=\displaystyle -\frac{\partial\ln(Z_\lambda)}{\partial\beta},\\
&=\frac{\beta   \left(\beta ^2 \lambda +16 \pi \right)}{128 \pi ^2} .
\end{align}

\begin{figure}[!ht]
\begin{tabbing}
\hspace{-1.5cm}
    \centering
	\includegraphics[width=0.55\linewidth]{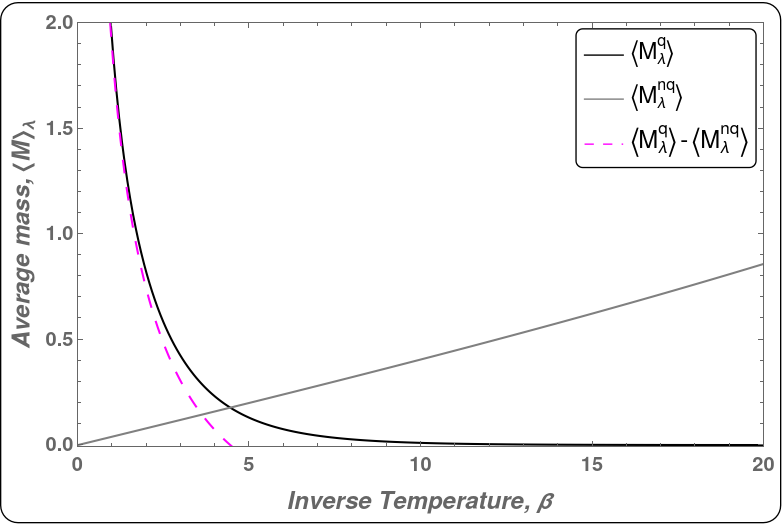}
    \hspace{0.4cm}
        \includegraphics[width=0.54\linewidth]{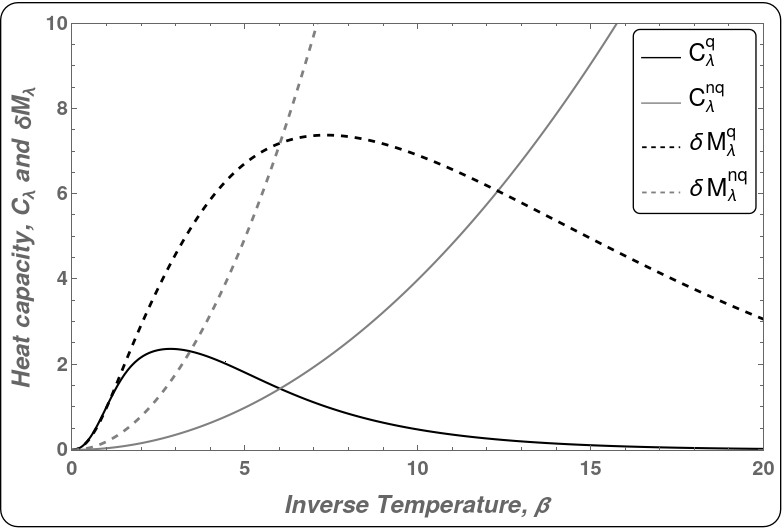}
\end{tabbing}
\vspace{-0.6cm}
	\caption{\it \footnotesize \textbf{\textit{Left panel:}} Quantized (black solid curve) and unquantized (gray solid curve) average mass of Schwarzschild R\'enyi-flat black hole in the canonical ensemble. The dashed curve (Magenta) represents the difference of the two averages. \textbf{\textit{Right panel:}} Heat capacity of the quantized (black solid line) and the unquantized R\'enyi black hole(gray solid line). Dashed curves in the two cases indicate black hole mass uncertainties. We fixed $\lambda=0.001$ and $\alpha=8$. }
	\label{fig:mass}
\end{figure}
Fig.\ref{fig:mass} illustrates the variations of the quantized and unquantized average masses of the R\'enyi-Schwarzschild black hole. The two masses exhibit distinct trends at low and high temperatures. At low temperature, while thermal radiation prevails for the quantized black hole state and on average no black hole exists, the unquantized mass scales polynomially with inverse temperature, which indicates simply that large black holes have smaller temperatures. At high temperature, the unquantized average goes to zero, the quantized mass grows significantly, yielding a dominant black hole phase.

\paragraph{} It is well known that heat capacity informs about the number of degrees of freedom of a thermodynamic system. Besides, heat capacity controls energy fluctuations around its mean value. One computes the heat capacity $C_\lambda$ of the R\'enyi-flat black hole through the partition functions, Eqs\eqref{eq_21} and \eqref{eq_21q} such as,
\begin{eqnarray}
    C_\lambda=\beta^2\frac{\partial^2 \ln(Z_\lambda)}{\partial \beta^2}.
\end{eqnarray}
The uncertainties in the mass of the black hole due to thermal fluctuations are measured in terms of the heat capacity,
\begin{equation}
    \delta M_\lambda=\beta \sqrt{C_\lambda}.
\end{equation}
On the right panel of Fig.\ref{fig:mass}, we plotted the heat capacity of the quantized (black line) and unquantized (gray line) of the uncharged R\'enyi-flat black hole as a function of inverse temperature.

\paragraph{} One can compute the R\'enyi density of states of the black hole, $g_\lambda(M)$, in the limit of small $\lambda$ through an \textit{inverse Laplace transform} of Eq.\eqref{eq_21} given by the Mellin's inverse formula\cite{Braden1987dos},
\begin{equation}
g_\lambda(M)=\displaystyle\underset{T\rightarrow\infty}{lim}\:\frac{1}{2\pi i}\int_{t-iT}^{t+iT}Z_\lambda^{(nq)}(\beta)e^{\beta M}d\beta,
\end{equation}
where the integration is done in the complex plane along the vertical line $Re(\beta)=t$ and $t$ is a real constant ensuring convergence\footnote[6]{$Re(.)$ is the real part function. Generally, $t$ is taken to be greater than all singularities of the integrand. for our case any choice of $t$ is valid.}. Carrying the integration one obtains,
\begin{equation}
g_\lambda(M) =\displaystyle 2 \,e^{4\pi M^2} \sum_{k=0}^\infty \frac{(-1)^k \lambda^k}{32^k k!} H_{4k}(2\sqrt{\pi} M),
\end{equation}
where $H_{4k}(x)$ are the well-known \textit{Hermite polynomials} of degree $4k$, $k\in \mathbb{N}$. To first order in $\lambda$ one gets ($k\leq1$),

\begin{equation}\label{eq_dos}
g_\lambda(M) =\displaystyle \underbrace{e^{4\pi M^2}}_{\text{Gaussian term}} \left[ 2 - \frac{\lambda}{16} \underbrace{\left(256\pi^2 M^4 - 96\pi M^2 + 12\right)}_{H_4(2\sqrt{\pi}M)} \right] + \mathcal{O}(\lambda^2),
\end{equation}

\begin{figure}[!ht]
	\centering
	\includegraphics[width=0.75\linewidth]{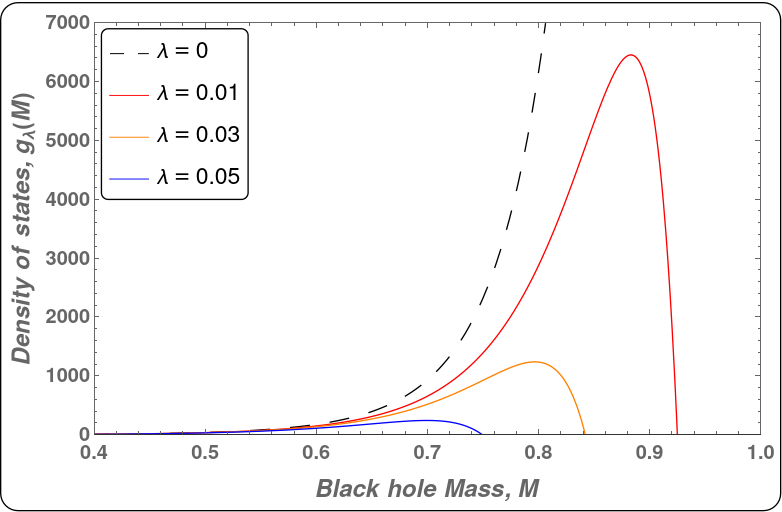}
	\caption{\it \footnotesize Density of states of the R\'enyi-Schwarzschild black hole in the canonical ensemble. The black dashed curve indicates a non-physical divergence at large mass in the absence of nonextensivity.}
	\label{fig:dos}
\end{figure}

The density of state Eq.\eqref{eq_dos} contains an expected Gaussian term expressing the relation, $\displaystyle g\propto e^{S}$ between the density of states and the black hole entropy, this comes from the Bekenstein-Hawking entropy verifying $\displaystyle S\propto A\propto M^2$. A notable feature of black holes is the large exponential increase of the density of states\cite{wald2001thermodynamics}, which permits high production rates in the Planckian regime compared with other final states. The nonextensivity dependent factor introduces a suppression for large ($M^4$ term) and small ($M^0$ constant term) masses but a relative enhancement for intermediate masses ($M^2$ term). Similar computations of the density of states of an asymptotically flat Schwarzschild black hole in different setups\cite{Andr2020dos,Salasnich2023dos,Rosabal2020dos} reproduce the same behavior.

Fig.\ref{fig:dos} portrays the modified density of states. It is evident that nonextensivity effects lower the density of states overall, but at higher masses, where the Schwarzschild black hole typically exhibits a non-physical divergent density of states, as shown by the black dashed curve, the behavior is significantly different.

%plot the density and comment  then comment the suppression of massive state M^4 and enhancing the intermidiate to small states M^2 term. also comment on the limit of validity of this expression (a condition on beta or lam). the constante term is bit strange also analyse its influence. may also the factor 2 at the begining

\paragraph{} In this final section we analyze a novel thermodynamic duality between seemingly different black hole systems, the R\'enyian asymptotically-flat black hole and the Boltzmannian asymptotically-AdS black hole.

\section{Duality of Charged-AdS Gibbs-Boltzmann vs Charged-flat R\'enyi black holes: A conformal transformation }
\label{sect4}
\begin{figure}[!ht]
 \vspace{0cm}
		\centering
		\begin{tabbing}
		\hspace{-1.6cm}
		\includegraphics[scale=0.45]{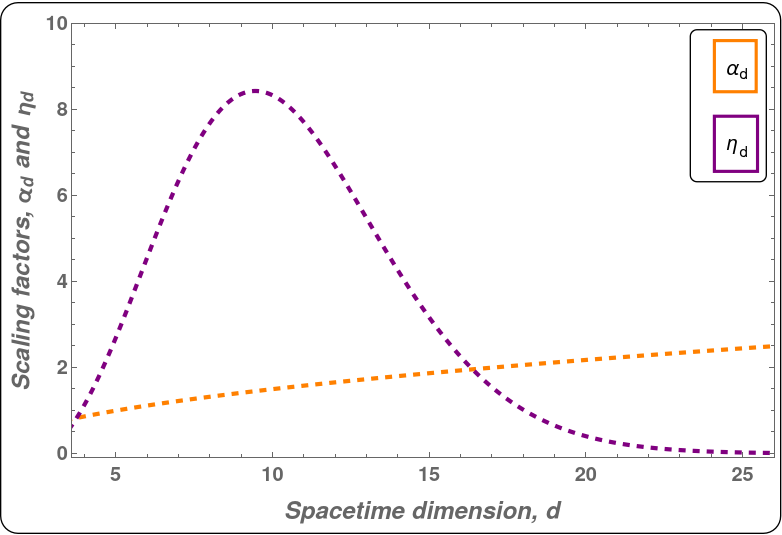}
		\hspace{0.1cm}
		\includegraphics[scale=0.45]{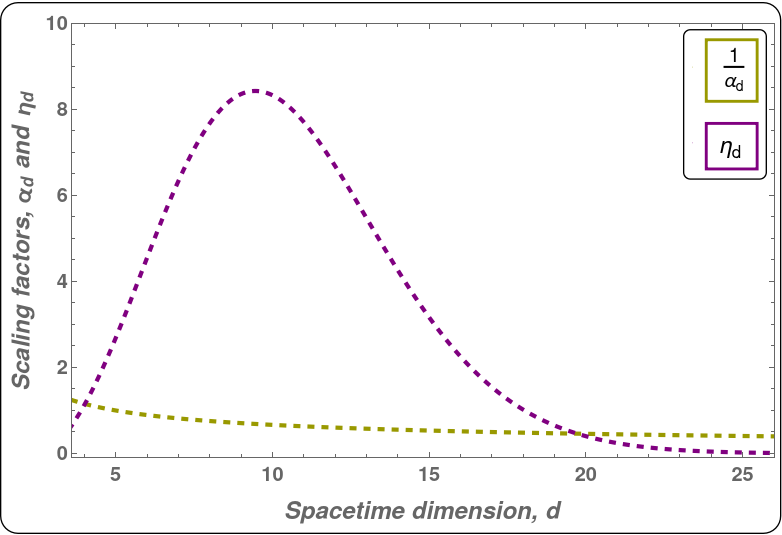}
		\end{tabbing}
		\vspace{-0.6cm}
		\caption{{\it\footnotesize Variations of the scaling factors with spacetime dimensions. }}
		\label{fig:scaling}
	\end{figure}

\paragraph{}A close inspection reveals a simple thermodynamic duality/equivalence between the charged-AdS Gibbs-Boltzmann (\textit{GB-AdS}) and the charged-flat R\'enyi (\textit{R\'enyi-flat}) black holes. In the present section, we provide an analysis and a mathematical formalism of such duality. An immediate consequence of this duality is the possibility to directly translate the microstructure of one system to the other, that is, any microscopic description available in one black hole, say \textit{GB-AdS}, can be easily pushed forward to the \textit{R\'enyi-flat} black hole and vice versa.
We begin from their respective equations of state in four-dimensional spacetime, which read,
\begin{equation}\label{eq41w}
P=T\rho-\frac{\rho^2}{2 \pi}+\frac{2 Q^2\rho^4}{\pi},
\end{equation}
\begin{equation}\label{eq42w}
P_R=  T_R\rho_R  - \frac{2\rho_R}{3 \pi } + \frac{128 Q_R^{2}\rho_R^4}{27 \pi}.
\end{equation}
 Where, we replaced the specific volumes $(v,\,v_R)$ by the volume densities $(\rho,\,\rho_R)$, respectively. A close inspection of the equations of state for the two black holes, Eqs.\eqref{eq41w} and \eqref{eq42w}, reveals the following observation: Given a Gibbs-Boltzmann-AdS black hole with state variables
 $
     \displaystyle\left(P,T,\rho,Q\right).
$
 There is a thermodynamic equivalent R\'enyi-flat black hole with the state variables
 $
      \displaystyle\left(P, \frac{T}{\xi},\xi\rho,\xi Q\right)
$
 where $\displaystyle\xi=\sqrt{\frac{3}{4}}$, that is
\begin{equation}\label{equiv}
\text{\textit{GB-AdS}}:\left(P,T,\rho,Q\right)\longleftrightarrow \text{\textit{R\'enyi-flat}}:\left(P, \frac{T}{\xi},\xi\rho,\xi Q\right).
\end{equation}
In other words, starting from the R\'enyi equation of state Eq.\eqref{eq42w}, and performing the following scaling, $T\rightarrow T/\xi$, $\rho\rightarrow \xi\rho$ and $Q\rightarrow \xi Q$, we recover the AdS equation of state, Eq.\eqref{eq41w}. The equivalence, Eq.\eqref{equiv}, is also scalable, meaning that a multiplication of both sides by a factor preserves the duality. In particular, a multiplication by $\xi$ gives,
\begin{equation}\label{eq107r}
\text{\textit{GB-AdS}}:(\xi P, \xi T,\xi \rho, \xi Q)\longleftrightarrow \text{\textit{R\'enyi-flat}}:(\xi P, T,\xi^2\rho,\xi^2 Q)
\end{equation}
Thus, a one-to-one correspondence between the $4d$ GB-AdS and R\'enyi-flat black holes holds. A generalization of this equivalence, Eq\eqref{equiv}, to higher dimensions is straightforward. In $d$-dimensional spacetime, the equations of states generalizes to\cite{Belhaj2012uiu},
\begin{spreadlines}{0.9em}
\begin{align}
    % &P=\displaystyle \frac{(d-2)T}{4r_h}-\frac{(d-2)(d-3)}{16 \pi r_h^2}+\frac{(d-2)(d-3)Q^2}{16\pi r_h^{2d-4}}\\
    &P=\displaystyle T\rho-\frac{(d-3)\rho^2}{\pi  (d-2)}+\frac{16^{d-3} (d-3)(d-2)^{5-2 d} Q^2\rho^{2 d-4}}{\pi }\label{eq_ads},\\ 
    % &P_R=\displaystyle \frac{(d-1) (d-2) T_R}{16  r_h}-\frac{(d-1)(d-2) \left(d-3\right) }{64 \pi  r_h^2}+\frac{(d-1)\Gamma \left(\frac{d-1}{2}\right)^2 Q^2 }{8\pi^{d-2}r_h^{2 d-4}}
    &P_R=
        \displaystyle T_R\rho_R -\frac{4 (d-3)\rho_R^2}{\pi  (d-1)(d-2)}+\frac{2^{8 d-19} (d-2)^{4-2d} (d-1)^{5-2d} \Gamma \left(\frac{d-1}{2}\right)^2 Q^2 \rho_R^{2d-4}}{\pi ^{d-2}} \label{eq_renyi},
\end{align}
\end{spreadlines}

where the specific volumes in $d$-dimensions, $\rho$ and $\rho_R$, are defined as
\begin{equation}
    \rho=\frac{d-2}{4\,l_p^{d-2} r_h}\quad \text{and,}\quad \rho_R=\frac{(d-1)(d-2)}{16\,l_p^{d-2} r_h}.
\end{equation}
Where Planck's length $l_p$ is made manifest for clarity. We obtain the transformation,
\begin{equation}\label{eq_trans}
    \text{\textit{GB-AdS}}:(P, T, \rho, Q)\longleftrightarrow \text{\textit{R\'enyi-flat}}:\left(P,\frac{T}{\alpha_d},\alpha_d\rho,\eta_d Q\right),
\end{equation}
here the parameters $\alpha_d$ and $\eta_d$ are given by
\begin{eqnarray}
  &\alpha_d=\displaystyle \sqrt{\frac{d-1}{4}}. \\
   & \eta_d=\displaystyle \sqrt{\frac{(d-2)(d-3)}{2(d-1)}}\frac{\pi ^{\frac{d-3}{2}}}{ \Gamma \left(\frac{d-1}{2}\right)}.
\end{eqnarray}
The Variations of the scaling factors in terms of spacetime dimensions are depicted in Fig.\ref{fig:scaling}
It is obvious that such a transformation is \textit{invertible}. Given a R\'enyi-flat black hole, one finds a unique GB-AdS black hole such that
\begin{equation}
  \text{\textit{R\'enyi-flat}}: \left(P_R,T_R,\rho_R,Q_R\right) \longleftrightarrow\text{\textit{GB-AdS}}:\left(P_R, \alpha_d T_R, \frac{\rho_R}{\alpha_d}, \frac{Q_R}{\eta_d}\right),
\end{equation} 
It is remarkable that the scaling factors $\alpha_d$ and $\eta_d$ are equal only for $d=4$ where they equal $\frac{\sqrt{3}}{2}$. As depicted in Fig.\ref{fig:scaling}, they approach equality around $d=20$; however, not for an integer value of the spacetime dimension. The factor $\alpha_d^{1}$ scales the temperature and the volume density, while $\eta_d$ scales the electric charge. From the right panel of Fig.\ref{fig:scaling}, we see that the R\'enyi dual/equivalent has larger temperature and smaller volume density for $d\geq 6$ but acquires smaller temperature and larger volume density for $d=4 \text{ and } 5$ with maximum attained at $d=4$. The electric charge, on the other hand, has smaller values up to $d=18$ and a maximum value $\frac{\pi ^3}{4} \sqrt{\frac{7}{6}}$ reached at $d=9$.

In $d=5$, we find $\alpha_5=1$ and $\displaystyle\eta_5=\frac{2}{\sqrt{3}\pi}$. This is interesting in the context of a proposed reformulation of the $AdS_5/CFT_4$ correspondence in terms of $\text{R\'enyi}_5/CFT_4$\cite{Barzi:2024vvo} as it implies a complete equality of state parameters for the uncharged black holes such that,

\begin{equation}
     \text{\textit{GB-AdS$_5$}}:(P, T, \rho)\longleftrightarrow \text{\textit{R\'enyi$_5$-flat}}:\left(P,T,\rho\right).
\end{equation}
In other words, only in $5$-dimensional spacetime that the two uncharged black hole fluids, the $GB\text{-}AdS$ and the $R\text{\'e}nyi\text{-}flat$, behave identically. Requiring the same energy content for the two black holes, $M_R=M$, induces a relation between their numbers of microstates; From the Smarr formulas in $d=5$
\begin{align}
   & M=\frac{3}{2}TS-PV,\\
   & M_R=\frac{3}{2}T_R S_R-\frac{3}{2}PV,
\end{align}
where the thermodynamic volumes, $V_R=V$, are also equal by virtue of the equality of the specific volumes, $\rho_R=\rho$, in this particular case. One therefore obtains
\begin{eqnarray}
    S_R=S+\frac{PV}{3 T}.
\end{eqnarray}
Generalizing to all spacetime dimensions, one gets
\begin{align}
 M=\frac{d-2}{d-3}TS-\frac{2}{d-3}PV,\quad and& \quad  M_R=\frac{d-2}{d-3}T_R S_R-\frac{d-2}{d-3}P_RV_R.\label{eq_smarr}\\
 V_R=\alpha_d^{d-1}V, \quad& P_R=P \quad and\quad T_R=\frac{T}{\alpha_d}.
\end{align}
Now, imposing mass equality yields,
\begin{equation}\label{eq_entropy_dual1}
    S_R=\alpha_d S+\frac{PV}{T}\left(\alpha_d^{d}-\frac{2\alpha_d}{d-2}\right).
\end{equation}
A curious numerical correspondence arises where the scaling factor squared in $4$-dimensional spacetime, $\alpha_4^2=\eta_4^2=\frac{3}{4}$, are equal to the ratio of the pressure of the two sides of $AdS_5/CFT_4$ duality when considered at finite temperature. Specifically, the ratios of the entropy and energy densities as well as of pressures for the weak and strong limits  of the 't Hooft coupling, $\lambda_t$, in the $CFT_4$ reads \cite{Witten1998,Berman2000}
\begin{eqnarray}
    \frac{P_{\lambda_t\rightarrow\infty}}{P_{\lambda_t\rightarrow0}}=\frac{s_{\lambda_t\rightarrow\infty}}{s_{\lambda_t\rightarrow0}}=\frac{\epsilon_{\lambda_t\rightarrow\infty}}{\epsilon_{\lambda_t\rightarrow0}}=\frac{3}{4}.
\end{eqnarray}
This numerical coincidence may hint at a connection between the $AdS$/\textit{R\'enyi} equivalence and the \textit{Weak/Strong} coupling limits of the dual $CFT$; Using the transformation, Eq.\eqref{eq_trans}, it is possible to deduce new transformations by first scaling the volume density in Eq.\eqref{eq_ads} such that,
\begin{align}\label{eq_131w}
      &\text{\textit{GB-AdS}}:(P, T, \zeta \rho, Q)\longleftrightarrow \text{\textit{GB-AdS}}:\left(\frac{P}{\zeta^2},\frac{T}{\zeta}, \rho, \zeta^{d-1}Q\right).
\end{align}
Applying the transformation for the scaled volume density gives,
\begin{align}\label{eq_132w}
    \text{\textit{GB-AdS}}:(P, T, \zeta\, \rho, Q)\longleftrightarrow \text{\textit{R\'enyi-flat}}:\left(P,\frac{T}{\alpha_d},\alpha_d\zeta\rho,\eta_d Q\right).
    \end{align}
   From Eqs.\eqref{eq_131w} and \eqref{eq_132w}, we deduce that, 
    \begin{align}\label{eq_133w}
     \text{\textit{GB-AdS}}:\left(\frac{P}{\zeta^2},\frac{T}{\zeta}, \rho, \zeta^{d-1}Q\right)\longleftrightarrow \text{\textit{R\'enyi-flat}}:\left(P,\frac{T}{\alpha_d},\alpha_d\zeta\rho,\eta_d Q\right).
\end{align}
Finally, choosing $\zeta=\alpha_d^{-1}$ in Eq.\eqref{eq_133w} produces the transformation
\begin{align}
   \implies \text{\textit{GB-AdS}}:\left(\alpha_d^2 P,\alpha_d T, \rho, \frac{Q}{\alpha_d^{d-1}}\right)\longleftrightarrow \text{\textit{R\'enyi-flat}}:\left(P,\frac{T}{\alpha_d},\rho,\eta_d Q\right),
\end{align}
where the black holes have the same volume density. If instead we take $\zeta=\displaystyle\alpha_d$ in Eq.\eqref{eq_133w}, one gets
\begin{align}\label{eq_T_inv}
\text{\textit{GB-AdS}}:\left(\frac{P}{\alpha_d^2},\frac{ T}{\alpha_d}, \rho, \alpha_d^{d-1}Q\right)\longleftrightarrow \text{\textit{R\'enyi-flat}}:\left(P,\frac{T}{\alpha_d},\alpha_d^2\rho,\eta_d Q\right).
\end{align}
Then multiplying both sides by an overall factor $\alpha_d$,
\begin{align}
\text{\textit{GB-AdS}}:\left(\frac{P}{\alpha_d},T, \alpha_d \rho, \alpha_d^{d}Q\right)\longleftrightarrow \text{\textit{R\'enyi-flat}}:\left(\alpha_d P,T,\alpha_d^3\rho,\alpha_d\,\eta_d Q\right).
\end{align}
In this form, the two black holes have the same thermal state, and the scaling $\alpha_d^2$ between the pressures is made manifest. In \textit{$4$-dimensional} spacetime, we have
\begin{equation}
    \frac{P_R}{P_{AdS}}=\frac{P_{\lambda_t\rightarrow\infty}}{P_{\lambda_t\rightarrow0}}=\alpha_4^2=\frac{3}{4}.
\end{equation}
Now assuming equal mass $M_R=M$, the Smarr formulas, Eqs.\eqref{eq_smarr}, give a relation between the entropies of the two uncharged black holes as,
\begin{equation}\label{eq_entropy_dual2}
    S_R=S-\frac{PV}{T}\left(\frac{\alpha_d^{d-2}}{d-2}-\frac{1}{\alpha_d^{d-2}}\right)
\end{equation}
One notices that in this case the entropy of the dual R\'enyi-flat black hole is smaller than the entropy of the dual GB-AdS black hole. Eqs.\eqref{eq_entropy_dual2} and \eqref{eq_entropy_dual1} can be put in a multiplicative form by considering the number of microstates $\displaystyle\Omega=e^S$,
\begin{equation}
  \displaystyle  \Omega_R=\exp\left(\gamma_d \frac{PT}{V}\right)\,\displaystyle\Omega^{\nu_d},
\end{equation}
where  $\Omega$ and $\Omega_R$ are number of microstates of the GB-AdS and R\'enyi-flat black holes, respectively and $\gamma_d$ and $\nu_d$ are $d$-dependent factors. From the transformation defined by Eqs.\eqref{eq_T_inv} and \eqref{eq_entropy_dual2}, that leave the temperature invariant, one has $\displaystyle\nu_d=1$ and $\displaystyle\gamma_d=\frac{1}{\alpha_d^{d-2}}-\frac{\alpha_d^{d-2}}{d-2}$. We note $\displaystyle N_0=\frac{PT}{V}$ the number of "molecules" of the ideal gas at the same pressure $P$, volume $V$, and temperature $T$ as the GB-AdS black hole fluid. Thus, we have the conformal transformation,
\begin{equation}
     \displaystyle  \Omega_R=\exp\left(\gamma_d N_0\right)\,\displaystyle\Omega^{\nu_d}.
\end{equation}
\paragraph{}We demonstrated the existence of a duality/conformal maps between the charged and uncharged \textit{GB-AdS} and \textit{R\'enyi-flat} black holes. By requiring invariant pressure, volume density, temperature, or electric charge and imposing invariant energy content, one obtains different transformations. The conformality of the transformation stems precisely from such a requirement in analogy with the conformal bootstrap approach, where in that case the invariance of the correlation functions is required. Furthermore, their phase profiles, critical points, dynamical and chaotic behaviors, and other thermodynamic properties should all be expected to be dual to each other, since this duality is based on the equation of states of the two black holes. To conclude, we present a mathematical formalization of this duality by introducing the following mathematical notations
\begin{itemize}
    \item Let $\mathbb{E}_{AdS}$ and $\mathbb{E}_R$ be the spaces of equilibrium states of the \textit{GB-AdS} and \textit{R\'enyi-flat} black holes, respectively. They are defined by the algebraic equations,
    \begin{spreadlines}{0.9em}
    	\begin{align}
    	&P-\displaystyle T\rho+\frac{(d-3)}{\pi  (d-2)}\rho^2-\frac{16^{d-3} (d-3)(d-2)^{5-2 d} }{\pi }Q^2\rho^{2 d-4}=0\label{eq_ads2}\\ 
    	&P_R-
    	\displaystyle T_R\rho_R +\frac{4 (d-3)}{\pi  (d-1)(d-2)}\rho_R^2-\frac{2^{8 d-19} (d-2)^{4-2d} (d-1)^{5-2d} \Gamma \left(\frac{d-1}{2}\right)^2}{\pi ^{d-2}} Q_R^2 \rho_R^{2d-4}=0 \label{eq_renyi2},
    	\end{align}
    \end{spreadlines}
    where $d$ is positive a integer $d>3$. Thus $\mathbb{E}_{AdS}$ and $\mathbb{E}_R$ are \textit{algebraic manifolds} on the field of real numbers, also called \textit{Nash manifolds}.
    
    \item Let $B_{AdS}$ and $B_R$ be the dual \textit{GB-AdS}  and \textit{R\'enyi-flat} black holes, respectively characterized by their states, $\Sigma_{AdS}(P,T,\rho,Q)$ and $\Sigma_R(P_R,T_R,\rho_R,Q_R)$. $\Sigma_{AdS}$  and $\Sigma_R$ are points in $\mathbb{E}_{AdS}$ and $\mathbb{E}_R$, respectively.
    \item We define the map,
    \begin{align}
    \begin{split}
 \pi_{Y}:\mathbb{E}_{AdS}&\longrightarrow\mathbb{E}_R\\
\left(B_{AdS},\Sigma_{AdS}\right)&\longrightarrow\left(B_R,\Sigma_R\right)
   \end{split} 
   \end{align} 
    as a conformal map that leaves the state variable $Y$ invariant. Where $Y\in\{P,\,T,\,\rho,Q\}$. There are four such maps
    \begin{align}
       \begin{split}
\displaystyle &\pi_{P}:\left(\mathbf{P},T,\rho,Q\right)\longrightarrow\left(\mathbf{P},\frac{T}{\alpha_d},\alpha_d\rho,\eta_d Q\right),\\
&\pi_T:\displaystyle\left(P,\mathbf{T}, \rho, Q\right)\longrightarrow \left(\alpha_d^2 P,\mathbf{T},\alpha_d^2\rho,\frac{\,\eta_d Q}{\alpha_d^{d-1}}\right),\\
&\displaystyle\pi_\rho:\left( P,T, \bm{\rho}, Q\right)\longrightarrow \left(\frac{P}{\alpha_d^2},\frac{T}{\alpha_d^2} ,\bm{\rho}, \alpha_d^{d-1}\eta_dQ\right),\\
&\pi_Q:\displaystyle\left(P,T, \rho, \mathbf{Q}\right)\longrightarrow \left(\eta_d^{\frac{2}{d-2}}P,\frac{\eta_d^{\frac{1}{d-1}}}{\alpha_d}T ,\alpha_d\eta_d^{\frac{1}{d-1}}\rho,\mathbf{Q}\right).
 \end{split}
    \end{align}
    For brevity, we note $\sigma_{YY^{'}}$ \textit{the scaling factor multiplying the state variable $Y^{'}$ in the map $\pi_Y$}, e.g. one can check that $\sigma_{PQ}=\eta_d$ and $\sigma_{\rho T}=\alpha_d^{2}$. By definition, one has $\sigma_{YY}=1$, $\forall\,Y\in \{P,\,T,\,\rho,\,Q\}$. We note also that $\sigma_{YY^{'}}$ are non-singular.
    \item $\pi_{Y}$ are \textit{linear} under multiplication by a nonzero real scalar,
    \begin{equation}\label{eq_148ee}
        \forall\,Y, \pi_{Y}(\mu\,\Sigma_{AdS})=\mu\,\pi_{Y}(\Sigma_{AdS}).
    \end{equation}
    
    \item The map $\pi_Y$ is \textit{invertible}, we note $\left(\pi_{Y}\right)^{-1}\equiv\overline{\pi}_Y:\mathbb{E}_R\longrightarrow\mathbb{E}_{AdS}$ its inverse and $\overline{\sigma}_{YY^{'}}$ the corresponding scaling factors. It is clear that $\overline{\sigma}_{YY^{'}}=\sigma^{-1}_{YY^{'}}$.
    \item We define the composition maps
    \begin{itemize}
        \item $\Pi_{YY^{'}}:\mathbb{E}_{AdS}\longrightarrow\mathbb{E}_{AdS}$ such as,
    \begin{equation}
       \Pi_{YY^{'}}=\overline{\pi}_{Y^{'}}\circ\pi_Y 
    \end{equation}
    \item$\overline{\Pi}_{YY^{'}}:\mathbb{E}_R\longrightarrow\mathbb{E}_R$ such as,
    \begin{equation}
      \overline{\Pi}_{YY^{'}}=\pi_{Y^{'}}\circ\overline{\pi}_Y 
    \end{equation}
   
        \end{itemize}
    
\item Each of the sets of transformations, $\Pi_{AdS}=\{\Pi_{YY^{'}}\}$ and $\overline{\Pi}_R=\{\Pi_{YY^{'}}\}$ have an \textit{algebraic group structure} under the associative composition operator $\circ$.  $\Pi_{YY}=id$ and $\overline{\Pi}_{YY}=\overline{id}$ are the identity elements of the groups $\Pi_{AdS}$ and $\overline{\Pi}_R$, respectively.

\item  $\Pi_{YY^{'}}$ and $\overline{\Pi}_{YY^{'}}$ are linear maps under scalar multiplication by virtue of the linearity of $\pi_Y$, Eq.\eqref{eq_148ee}.
 \item It is easy to show that, 
 \begin{equation}
\displaystyle\left(\Pi_{YY^{'}}\right)^{-1}\equiv\overline{\Pi_{YY^{'}}}=\overline{\Pi}_{Y^{'}Y}.
 \end{equation}
 \item The commutative diagram constructed such as,
\begin{equation}
\begin{tikzcd}[column sep=huge,row sep=huge]
\mathbb{E}_{AdS} \arrow[r, "\displaystyle\pi_{Y_1}"] \arrow[d,"\displaystyle\Pi_{2}"] \arrow[rd,"\displaystyle\pi_{Y_3}"]
& \mathbb{E}_R\arrow[d, "\displaystyle\overline{\Pi}_{1}"] \\
\mathbb{E}_{AdS} \arrow[r, black, "\displaystyle\pi_{Y_2}" black]
& [black] \mathbb{E}_R
\end{tikzcd}  
\end{equation}
furnishes the relations, $\pi_{Y_3}=\overline{\Pi}_1\circ\pi_{Y1}=\pi_{Y_2}\circ\Pi_2$.
\item Given that the transformations $\pi_{Y}$ are represented by diagonal matrices
\begin{equation}
\begin{pmatrix}
\sigma_{YP} & &    &\text{\huge0}\\
 & \sigma_{YT}  & &\\
&  & \sigma_{Y\rho} &\\
\text{\huge0}&& &\sigma_{YQ} 
\end{pmatrix}.
\end{equation}
The groups $\Pi_{AdS}$ and $\Pi_R$ are \textit{abelian}. That is, $\Pi_{1}\Pi_{2}=\Pi_{2}\Pi_{1}$. The group elements of $\Pi_{AdS}$ and $\Pi_R$ are also represented by diagonal matrices
\begin{equation}
\Pi_{YY^{'}}=
\begin{pmatrix}
\overline{\sigma}_{Y^{'}P}\,\sigma_{YP} & &    &\text{\huge0}\\
 & \overline{\sigma}_{Y^{'}T}\,\sigma_{YT}  & &\\
&  & \overline{\sigma}_{Y^{'}\rho}\, \sigma_{Y\rho} &\\
\text{\huge0}&& &\overline{\sigma}_{Y^{'}Q} \,\sigma_{YQ}
\end{pmatrix}.
\end{equation}
\item Through inspection, it is possible to derive new transformations such as, 
\begin{equation}
\displaystyle\overline{\sigma}_{Y^{'}Y_i}\,\sigma_{YY_i}\,\Pi_{YY^{'}}(\Sigma_{AdS}),\quad \forall \,Y,\,Y^{'} \text{and }Y_i
\end{equation} 
leaves invariant at least the state variable $Y_i$ of the \textit{GB-AdS} black hole. An interesting example is the transformation, $\alpha_d^{-1}\,\Pi_{PT}(\Sigma_{AdS})$, which keeps invariant the temperature $T$ and the volume density $\rho$ of the black hole.

\end{itemize}

It is interesting to investigate the actions of the groups $\Pi_{AdS}$ and $\Pi_{R}$ on the manifolds $\mathbb{E}_{AdS}$ and $\mathbb{E}_{R}$, respectively. Specifically, whether each action generates a foliation of the manifold into orbits or equivalence classes of black holes. Subsequent developments on this matter and related ones will constitute the focus of future work.

\section{Conclusion}\label{sect5}
In this paper, we have explored the thermodynamics and statistical mechanics of the asymptotically flat black hole within the Rényi entropy framework, with a particular emphasis on its quantized structure. Starting from the quantized area spectrum, we derived the Rényi entropy and showed that the number of black hole microstates, and consequently the number of information bits, is significantly affected by the nonextensive parameter $\lambda$. Notably, each bit encodes occupies a number of Planck area units which we called occupancy, leading to a deeper microscopic interpretation of black hole entropy.

Moreover, We investigated the thermodynamic implications of this framework, including first-order phase transitions between small and large black holes. The latent heat was shown to emerge from a discontinuous jump in the number of area quanta, and its expression was derived both via the Clapeyron relation and directly from the entropy difference. We also constructed the partition function and probability distribution for the quantized Rényi-Schwarzschild black hole and compared it with the continuous case.  We showed that the density of states gets a correction from nonextensivity effects inducing a suppression for large black holes and eliminates the divergence predicted by Gibbs-Boltzmann statistics at large black hole masses. The results highlight the effect of nonextensivity at finite temperatures, where the distribution significantly deviates from its Boltzmann–Gibbs counterpart, especially at high energies.

Furthermore, we established a thermodynamic duality between the charged AdS black hole in Boltzmann–Gibbs statistics and the charged-flat black hole in Rényi statistics. This duality was expressed through a conformal transformation that maps thermodynamic variables between the two systems. We unveiled that this mapping generalizes to higher dimensions, with dimension-dependent scaling parameters $\alpha_d$ and $\eta_d$, which coincide only in four-dimensional spacetime with a particular correspondence for the uncharged black holes in five-dimensional spacetime. Moreover, we linked the number of microstates of the dual black holes through a simple conformal relation which invokes the great similitude these two systems display. Finally, This investigation gave rise to an interesting and rich mathematical structure from algebraic geometry underlying such duality whose deep consequences we attend to uncover.

Overall, our findings underscore the effectiveness of Rényi statistics in establishing a coherent nonextensive thermodynamic framework for black holes. This approach sheds new light on their microscopic structure, phase transitions, and statistical behavior. Moreover, the formalism reveals a compelling pathway to relate asymptotically flat and AdS black holes via entropy deformations and conformal mappings, suggesting the presence of a deeper underlying duality that warrants further investigation.

\section*{Acknowledgements}
H. El M would like to acknowledge networking
support of the COST Action 
 CA 22113 - Fundamental challenges in theoretical physics (Theory and Challenges), 
CA 21136 - Addressing observational tensions in cosmology with systematics and fundamental physics (CosmoVerse), and 
CA 23130 - Bridging high and low energies in search of quantum gravity (BridgeQG). He also thanks IOP for its support. This work was carried out under the project UIZ 2025 Scientific Research Projects.

%\newpage

%\appendix
%\numberwithin{equation}{section}
%\newpage

\bibliographystyle{unsrt}
\bibliography{microstructure} 

\begin{thebibliography}{10}

\bibitem{strominger1996microscopic}
Andrew Strominger and Cumrun Vafa.
\newblock Microscopic origin of the bekenstein-hawking entropy.
\newblock {\em Physics Letters B}, 379(1-4):99--104, 1996.

\bibitem{mathur2005fuzzball}
Samir~D Mathur.
\newblock The fuzzball proposal for black holes: an elementary review.
\newblock {\em Fortschritte der Physik}, 53(7):793--827, 2005.

\bibitem{maldacena1999large}
Juan~M Maldacena.
\newblock The large-n limit of superconformal field theories and supergravity.
\newblock {\em International Journal of Theoretical Physics}, 38:1113--1133, 1999.

\bibitem{hawking1976breakdown}
Stephen~W Hawking.
\newblock Breakdown of predictability in gravitational collapse.
\newblock {\em Physical Review D}, 14(10):2460, 1976.

\bibitem{bekenstein1973black}
Jacob~D Bekenstein.
\newblock Black holes and entropy.
\newblock {\em Physical Review D}, 7(8):2333, 1973.

\bibitem{hawking1975particle}
Stephen~W Hawking.
\newblock Particle creation by black holes.
\newblock {\em Communications in Mathematical Physics}, 43(3):199--220, 1975.

\bibitem{bardeen1973four}
James~M Bardeen, Brandon Carter, and Stephen~W Hawking.
\newblock The four laws of black hole mechanics.
\newblock {\em Communications in Mathematical Physics}, 31(2):161--170, 1973.

\bibitem{wald2001thermodynamics}
Robert~M Wald.
\newblock The thermodynamics of black holes.
\newblock {\em Living Reviews in Relativity}, 4(1):1--44, 2001.

\bibitem{tsallis1988possible}
Constantino Tsallis.
\newblock Possible generalization of boltzmann-gibbs statistics.
\newblock {\em Journal of Statistical Physics}, 52(1-2):479--487, 1988.

\bibitem{padmanabhan2010thermodynamical}
Thanu Padmanabhan.
\newblock Thermodynamical aspects of gravity: new insights.
\newblock {\em Reports on Progress in Physics}, 73(4):046901, 2010.

\bibitem{renyi1961measures}
Alfréd Rényi.
\newblock On measures of information and entropy.
\newblock {\em Proceedings of the Fourth Berkeley Symposium on Mathematical Statistics and Probability}, 1:547--561, 1961.

\bibitem{biro2013thermodynamics}
T~S Bir{\'o} and P~V{\'a}n.
\newblock Thermodynamics of composition in non-extensive systems.
\newblock {\em Physical Review E}, 87(4):042123, 2013.

\bibitem{czinner2016renyi}
Viktor~G Czinner and Hideo Iguchi.
\newblock Rényi entropy and the thermodynamic stability of black holes.
\newblock {\em Physics Letters B}, 752:306--310, 2016.

\bibitem{czinner2017generalized}
Viktor~G Czinner and Hideo Iguchi.
\newblock Generalized entropy and modified black hole thermodynamics.
\newblock {\em The European Physical Journal C}, 77:1--7, 2017.

\bibitem{Promsiri:2020jga}
Chatchai Promsiri, Ekapong Hirunsirisawat, and Watchara Liewrian.
\newblock {Thermodynamics and Van der Waals phase transition of charged black holes in flat spacetime via R\'enyi statistics}.
\newblock {\em Phys. Rev. D}, 102(6):064014, 2020.

\bibitem{promson2020renyi}
A~Promson and A~Chatrabhuti.
\newblock Rényi entropy and thermodynamic stability of de sitter black holes.
\newblock {\em The European Physical Journal C}, 80:1--13, 2020.

\bibitem{Barzi2022}
F.~Barzi and H.~El Moumni.
\newblock On {R}ényi universality formula of charged flat black holes from hawking-page phase transition.
\newblock {\em Physics Letters B}, 833:137378, 10 2022.

\bibitem{Barzi:2023mit}
F.~Barzi, H.~El~Moumni, and K.~Masmar.
\newblock {On some phase equilibrium features of charged black holes in flat spacetime via {R}\'enyi statistics}.
\newblock {\em Gen. Rel. Grav.}, 55(10):109, 2023.

\bibitem{Barzi:2024vvo}
F.~Barzi, H.~El~Moumni, and K.~Masmar.
\newblock {Nonextensive black hole thermodynamics from generalized Euclidean path integral and Wick{\textquoteright}s rotation}.
\newblock {\em Eur. Phys. J. C}, 85(1):61, 2025.
\newblock [Erratum: Eur.Phys.J.C 85, 211 (2025)].

\bibitem{Barzi2024yan}
F.~Barzi, H.~El Moumni, and K.~Masmar.
\newblock Thermal chaos of charged-flat black hole via {R}ényi formalism.
\newblock {\em Nuclear Physics B}, 1005:116606, 8 2024.

\bibitem{Barzi2024yin}
F.~Barzi, H.~El Moumni, and K.~Masmar.
\newblock {R}ényi topology of charged-flat black hole: Hawking-page and van-der-waals phase transitions.
\newblock {\em Journal of High Energy Astrophysics}, 42:63--86, 6 2024.

\bibitem{Dong2016re}
Xi~Dong, Aitor Lewkowycz, and Mukund Rangamani.
\newblock The rényi entropy of a theory with a gravity dual.
\newblock {\em Journal of High Energy Physics}, 2016(11):28, 2016.

\bibitem{Lewkowycz2013generalized}
Aitor Lewkowycz and Juan Maldacena.
\newblock Generalized gravitational entropy.
\newblock {\em Journal of High Energy Physics}, 2013(8):90, 2013.

\bibitem{wei2015insight}
Shao-Wen Wei, Yu-Xiao Liu, et~al.
\newblock Insight into the microscopic structure of an ads black hole from a thermodynamical phase transition.
\newblock {\em Physical review letters}, 115(11):111302, 2015.

\bibitem{KordZangeneh:2017lgs}
M.~Kord~Zangeneh, A.~Dehyadegari, A.~Sheykhi, and R.~B. Mann.
\newblock {Microscopic Origin of Black Hole Reentrant Phase Transitions}.
\newblock {\em Phys. Rev. D}, 97(8):084054, 2018.

\bibitem{base}
Dao-Quan Sun, Jian-Bo Deng, Ping Li, and Xian-Ru Hu.
\newblock {Insight into the Microscopic Structure of an AdS Black Hole from the Quantization}.
\newblock {\em Class. Quant. Grav.}, 37(1):015008, 2020.

\bibitem{Chabab:2017xdw}
M.~Chabab, H.~El~Moumni, S.~Iraoui, K.~Masmar, and S.~Zhizeh.
\newblock {More Insight into Microscopic Properties of RN-AdS Black Hole Surrounded by Quintessence via an Alternative Extended Phase Space}.
\newblock {\em Int. J. Geom. Meth. Mod. Phys.}, 15(10):1850171, 2018.

\bibitem{Miao:2017cyt}
Yan-Gang Miao and Zhen-Ming Xu.
\newblock {Microscopic structures and thermal stability of black holes conformally coupled to scalar fields in five dimensions}.
\newblock {\em Nucl. Phys. B}, 942:205--220, 2019.

\bibitem{David:2002wn}
Justin~R. David, Gautam Mandal, and Spenta~R. Wadia.
\newblock {Microscopic formulation of black holes in string theory}.
\newblock {\em Phys. Rept.}, 369:549--686, 2002.

\bibitem{Mirza:2007ev}
Behrouz Mirza and Mohammad Zamani-Nasab.
\newblock {Ruppeiner Geometry of RN Black Holes: Flat or Curved?}
\newblock {\em JHEP}, 06:059, 2007.

\bibitem{Niu:2011tb}
Chao Niu, Yu~Tian, and Xiao-Ning Wu.
\newblock {Critical Phenomena and Thermodynamic Geometry of RN-AdS Black Holes}.
\newblock {\em Phys. Rev. D}, 85:024017, 2012.

\bibitem{Dehyadegari:2016nkd}
Amin Dehyadegari, Ahmad Sheykhi, and Afshin Montakhab.
\newblock {Critical behavior and microscopic structure of charged AdS black holes via an alternative phase space}.
\newblock {\em Phys. Lett. B}, 768:235--240, 2017.

\bibitem{Emparan:2006it}
Roberto Emparan and Gary~T. Horowitz.
\newblock {Microstates of a Neutral Black Hole in M Theory}.
\newblock {\em Phys. Rev. Lett.}, 97:141601, 2006.

\bibitem{Horowitz:1996fn}
Gary~T. Horowitz and Andrew Strominger.
\newblock {Counting states of near extremal black holes}.
\newblock {\em Phys. Rev. Lett.}, 77:2368--2371, 1996.

\bibitem{Maldacena:1996gb}
Juan~Martin Maldacena and Andrew Strominger.
\newblock {Statistical entropy of four-dimensional extremal black holes}.
\newblock {\em Phys. Rev. Lett.}, 77:428--429, 1996.

\bibitem{Aounallah:2022rfo}
H.~Aounallah, H.~El~Moumni, J.~Khalloufi, and K.~Masmar.
\newblock {Insight into the microscopic structure of a quintessential black hole from the quantization concept}.
\newblock {\em Int. J. Mod. Phys. A}, 37(08):2250036, 2022.

\bibitem{Renyi1959aqqq}
A.~R{\'e}nyi.
\newblock On the dimension and entropy of probability distributions.
\newblock {\em Acta Mathematica Academiae Scientiarum Hungarica}, 10(1):193--215, 1959.

\bibitem{Czinner:2015eyk}
Viktor~G. Czinner and Hideo Iguchi.
\newblock {R{\'e}nyi Entropy and the Thermodynamic Stability of Black Holes}.
\newblock {\em Phys. Lett. B}, 752:306--310, 2016.

\bibitem{Tannukij:2020njz}
Lunchakorn Tannukij, Pitayuth Wongjun, Ekapong Hirunsirisawat, Tanapat Deesuwan, and Chatchai Promsiri.
\newblock {Thermodynamics and Phase Transition of Spherically Symmetric Black Hole in de Sitter Space from R\'enyi Statistics}.
\newblock {\em Eur. Phys. J. Plus}, 135(6):500, 2020.

\bibitem{dilaton}
Hasan El~Moumni, Karima Masmar, and Safaa Mazzou.
\newblock {Critical phenomena of charged dilatonic black holes through R\'enyi statistics approach}.
\newblock {\em Int. J. Mod. Phys. D}, 31(05):2250040, 2022.

\bibitem{Abeintro}
Sumiyoshi Abe.
\newblock General pseudoadditivity of composable entropy prescribed by the existence of equilibrium.
\newblock {\em Phys. Rev. E}, 63:061105, May 2001.

\bibitem{Rajagopal2014van}
Aruna Rajagopal, David Kubizňák, and Robert~B. Mann.
\newblock Van der waals black hole.
\newblock {\em Physics Letters B}, 737:277--279, 10 2014.

\bibitem{Ditta2023van}
Allah Ditta, Xia Tiecheng, Riasat Ali, and G.~Mustafa.
\newblock Thermal stability and tunneling radiation in van der waals black hole.
\newblock {\em Nuclear Physics B}, 994:116287, 9 2023.

\bibitem{Kaczmarek2024vann}
Adam~Z. Kaczmarek, Yassine Sekhmani, Dominik Szcz\c{e}\'sniak, and Javlon Rayimbaev.
\newblock The thermodynamics of the van der waals black hole within kaniadakis entropy.
\newblock {\em Entropy}, 26:1027, 11 2024.

\bibitem{Valero2025gup}
Ezequiel Valero, Hector Gisbert, and Victor Ilisie.
\newblock The generalized uncertainty principle. new bounds and trends.
\newblock 5 2025.

\bibitem{Bosso2023gup}
Pasquale Bosso, Giuseppe~Gaetano Luciano, Luciano Petruzziello, and Fabian Wagner.
\newblock 30 years in: Quo vadis generalized uncertainty principle?
\newblock {\em Classical and Quantum Gravity}, 40:195014, 9 2023.

\bibitem{Maggiore1993gup}
Michele Maggiore.
\newblock The algebraic structure of the generalized uncertainty principle.
\newblock {\em Physics Letters B}, 319:83--86, 12 1993.

\bibitem{Ong2025gup}
Yen~Chin Ong.
\newblock The case for black hole remnants: A review.
\newblock 1 2025.

\bibitem{Wang2024uiu}
Rui-Bo Wang, Shi-Jie Ma, Lei You, Yu-Cheng Tang, Yu-Hang Feng, Xian-Ru Hu, and Jian-Bo Deng.
\newblock Thermodynamics of ads-schwarzschild-like black hole in loop quantum gravity.
\newblock {\em The European Physical Journal C}, 84:1161, 11 2024.

\bibitem{Barbero1995uiu}
J.~Fernando~Barbero G.
\newblock Real ashtekar variables for lorentzian signature space-times.
\newblock {\em Physical Review D}, 51:5507--5510, 5 1995.

\bibitem{Ruppeiner1979eee}
George Ruppeiner.
\newblock Thermodynamics: A riemannian geometric model.
\newblock {\em Physical Review A}, 20:1608--1613, 10 1979.

\bibitem{Ruppeiner1995ttt}
George Ruppeiner.
\newblock Riemannian geometry in thermodynamic fluctuation theory.
\newblock {\em Reviews of Modern Physics}, 67:605--659, 7 1995.

\bibitem{Quevedo:2006xk}
Hernando Quevedo.
\newblock {Geometrothermodynamics}.
\newblock {\em J. Math. Phys.}, 48:013506, 2007.

\bibitem{Quevedo:2007mj}
Hernando Quevedo.
\newblock {Geometrothermodynamics of black holes}.
\newblock {\em Gen. Rel. Grav.}, 40:971--984, 2008.

\bibitem{Quevedo:2010tz}
Hernando Quevedo, Alberto Sanchez, Safia Taj, and Alejandro Vazquez.
\newblock {Phase transitions in geometrothermodynamics}.
\newblock {\em Gen. Rel. Grav.}, 43:1153--1165, 2011.

\bibitem{Weinhold1975geo}
F.~Weinhold.
\newblock Metric geometry of equilibrium thermodynamics.
\newblock {\em The Journal of Chemical Physics}, 63:2479--2483, 9 1975.

\bibitem{Hermann1973geo}
R.~Hermann.
\newblock Geometry, physics, and systems.
\newblock {\em M. Dekker}, page 304, 1973.

\bibitem{Mrugaa1978geo}
R.~MrugaŁa.
\newblock Geometrical formulation of equilibrium phenomenological thermodynamics.
\newblock {\em Reports on Mathematical Physics}, 14:419--427, 12 1978.

\bibitem{Quevedo2003geo}
H.~Quevedo and R.D. Zarate.
\newblock Differential geometry and thermodynamics.
\newblock 49:125--126, 2003.

\bibitem{Belhaj:2015uwa}
A.~Belhaj, M.~Chabab, H.~El~Moumni, K.~Masmar, and M.~B. Sedra.
\newblock {On Thermodynamics of AdS Black Holes in M-Theory}.
\newblock {\em Eur. Phys. J. C}, 76(2):73, 2016.

\bibitem{Chabab:2015ytz}
M.~Chabab, H.~El~Moumni, and K.~Masmar.
\newblock {On thermodynamics of charged AdS black holes in extended phases space via M2-branes background}.
\newblock {\em Eur. Phys. J. C}, 76(6):304, 2016.

\bibitem{Oshima1999geo}
Hiroshi Oshima, Tsunehiro Obata, and Hiroaki Hara.
\newblock Riemann scalar curvature of ideal quantum gases obeying gentile's statistics.
\newblock {\em Journal of Physics A: Mathematical and General}, 32:6373, 9 1999.

\bibitem{Ladino2025geo}
Jose~Miguel Ladino, Carlos~E. Romero-Figueroa, and Hernando Quevedo.
\newblock Phase transitions, shadows, and microstructure of kerr-anti-de sitter black holes from geometrothermodynamics.
\newblock {\em Nuclear Physics B}, 1018:117031, 9 2025.

\bibitem{Ruppeiner2008}
George Ruppeiner.
\newblock {Thermodynamic curvature and phase transitions in Kerr-Newman black holes}.
\newblock {\em Physical Review D - Particles, Fields, Gravitation and Cosmology}, 78(2):1--13, 2008.

\bibitem{Wei:2015iwa}
Shao-Wen Wei and Yu-Xiao Liu.
\newblock {Insight into the Microscopic Structure of an AdS Black Hole from a Thermodynamical Phase Transition}.
\newblock {\em Phys. Rev. Lett.}, 115(11):111302, 2015.
\newblock [Erratum: Phys.Rev.Lett. 116, 169903 (2016)].

\bibitem{patrick2017ob}
Patrick~Das Gupta and Eklavya Thareja.
\newblock Supermassive black holes from collapsing dark matter bose–einstein condensates.
\newblock {\em Classical and Quantum Gravity}, 34:035006, 2 2017.

\bibitem{Shemmer2004ob}
O.~Shemmer, H.~Netzer, R.~Maiolino, E.~Oliva, S.~Croom, E.~Corbett, and L.~di~Fabrizio.
\newblock Near‐infrared spectroscopy of high‐redshift active galactic nuclei. i. a metallicity–accretion rate relationship.
\newblock {\em The Astrophysical Journal}, 614:547--557, 10 2004.

\bibitem{Kozłowski2017ob}
Szymon Kozłowski.
\newblock Virial black hole mass estimates for 280,000 agns from the sdss broadband photometry and single-epoch spectra.
\newblock {\em The Astrophysical Journal Supplement Series}, 228:9, 1 2017.

\bibitem{Ge2019ob}
Xue Ge, Bi-Xuan Zhao, Wei-Hao Bian, and Green~Richard Frederick.
\newblock The blueshift of the c iv broad emission line in qsos.
\newblock {\em The Astronomical Journal}, 157:148, 3 2019.

\bibitem{Barzi2025info}
F~Barzi and K~Fethi.
\newblock Reformulation of classical thermodynamics from information theory.
\newblock {\em Physics Education}, 60:025501, 3 2025.

\bibitem{Braden1987dos}
Harry~W. Braden, Bernard~F. Whiting, and James~W. York.
\newblock Density of states for the gravitational field in black-hole topologies.
\newblock {\em Physical Review D}, 36:3614, 12 1987.

\bibitem{Andr2020dos}
Rui André and José~P.S. Lemos.
\newblock Thermodynamics of five-dimensional schwarzschild black holes in the canonical ensemble.
\newblock {\em Physical Review D}, 102:024006, 7 2020.

\bibitem{Salasnich2023dos}
Luca Salasnich.
\newblock Density of states for the unitary fermi gas and the schwarzschild black hole.
\newblock {\em Symmetry 2023, Vol. 15, Page 350}, 15:350, 1 2023.

\bibitem{Rosabal2020dos}
J.~A. Rosabal.
\newblock Schwarzschild black hole states and entropies on a nice slice.
\newblock {\em The European Physical Journal C 2020 80:12}, 80:1--18, 12 2020.

\bibitem{Belhaj2012uiu}
A.~Belhaj, M.~Chabab, H.~El Moumni, and M.~B. Sedra.
\newblock On thermodynamics of ads black holes in arbitrary dimensions.
\newblock {\em Chinese Physics Letters}, 29:100401, 10 2012.

\bibitem{Witten1998}
Edward Witten.
\newblock Anti de sitter space and holography.
\newblock {\em Adv.Theor.Math.Phys}, 2:253--291, 1998.

\bibitem{Berman2000}
David Berman.
\newblock Aspects of holography and rotating ads black holes.
\newblock In {\em Proceedings of Quantum aspects of gauge theories, supersymmetry and unification — PoS(tmr99)}, page~8. Sissa Medialab, 2 2000.

\end{thebibliography}
\end{document}